\documentclass[acmsmall,nonacm]{acmart}
\usepackage{stfloats}
\usepackage{longtable}
\usepackage{booktabs}
\usepackage{colortbl}
\usepackage{xcolor}
\usepackage{multirow}
\usepackage{xr-hyper}
\usepackage{tikz}
\usetikzlibrary{calc}
\usetikzlibrary{arrows.meta, positioning, fit, backgrounds, decorations.pathreplacing}

\newcommand{\imp}[1]{\cellcolor{green!15}{#1}}
\newcommand{\dgr}[1]{\cellcolor{red!15}{#1}}
\newcommand{\mis}[1]{\cellcolor{orange!20}{#1}}

\AtBeginDocument{%
  \providecommand\BibTeX{{%
    \normalfont B\kern-0.5em{\scshape i\kern-0.25em b}\kern-0.8em\TeX}}}

\acmVolume{0}
\acmNumber{CSCW}
\acmArticle{0}
\acmYear{2026}
\acmDOI{}

\begin{document}

\title{Fairness Theatre: Evaluating Post-Hoc Fairness Interventions in Vendor-Controlled Early Warning Systems}

\author{Kelly McConvey}
\affiliation{%
  \institution{University of Toronto}
  \city{Toronto}
  \state{Ontario}
  \country{Canada}
}
\email{kelly.mcconvey@mail.utoronto.ca}

\author{Angelina Zhai}
\affiliation{%
  \institution{Georgia Institute of Technology}
  \city{Atlanta}
  \state{Georgia}
  \country{USA}
}
\email{angelina.zhai@gatech.edu}

\author{Rebecca Li}
\affiliation{%
  \institution{University of Toronto}
  \city{Toronto}
  \state{Ontario}
  \country{Canada}
}
\email{brecca.li@mail.utoronto.ca}

\author{Shion Guha}
\affiliation{%
  \institution{University of Toronto}
  \city{Toronto}
  \state{Ontario}
  \country{Canada}
}
\email{shion.guha@utoronto.ca}

\begin{abstract}
Public institutions increasingly procure AI systems whose design they cannot inspect or change. In higher education, proprietary Early Warning Systems (EWS) leave colleges with few options beyond adjusting model outputs to address inequity. This raises a CSCW question: how is fairness work coordinated among vendors, institutions, advisors, and students with unequal power to change these systems? Using student records from a public college in Ontario, Canada, we evaluate six post-hoc fairness interventions on a research EWS under simulated procurement constraints. We compare fairness, accuracy, and demographic disparities, introducing error-type profiling to trace how interventions redistribute false positives and false negatives. Interventions redistributed disparities without consistently reducing them. Two implementations favored already-advantaged groups because they used group size to define disadvantage; small, marginalized groups remained poorly served. These findings show how procurement constraints and implementation choices shape the possibilities for fairness work. We call the resulting condition \emph{fairness theatre}: dashboard metrics converge while groups’ error burdens persist or worsen.
\end{abstract}



\received{March 2026}
\received[revised]{July 2026}
\received[revised]{September 2026}

\maketitle

\section{Introduction}
Early warning systems (EWS) make a simple promise in higher education: if institutions can identify students at high risk of not finishing their programs, they can step in early, improve outcomes, and keep more students enrolled. Critics have raised serious concerns about that promise. Likelihood of dropping out is not a reliable proxy for needing help \cite{andalibi_conceptualizing_2023}; there is limited evidence that the interventions triggered by these systems' predictions actually reduce attrition \cite{mcconvey_human-centered_2023}; and, perhaps most troubling, EWS may distribute resources unfairly across student populations \cite{baker_algorithmic_2022}.
When an EWS mis-allocates interventions, both students and institutions are impacted. Students in need of support slip through the cracks while students at low risk get flagged for attributes they cannot change, like age or postal code, and carry the stigma of having been singled out. The institution spends scarce advising resources on students it has misidentified, with no gain in retention or student experience. That scarcity also means institutions rarely build these systems themselves. As we describe in Section~\ref{sec:background}, EWS usually arrive through vendor procurement, which puts model design beyond the institution's reach and leaves adjusting the model's outputs as one of the few levers it has for impacting equity and fairness. 

This, in turn, becomes a coordination problem of the kind CSCW has long studied. Vendors, institutions, advisors, and students each hold part of what fairness requires, but they do not hold it equally. Vendors make the consequential design decisions: which features to use, how to define the outcome, what to optimize for. Institutions answer for whether the resulting classifications are equitable but cannot see or change how they were produced. Advisors act on scores they did not shape, and students live with errors they cannot see or contest. The contract establishes this distribution of control, granting the vendor authority over the model and leaving the institution to govern only its outputs. Fairness here is not a property of the model but something the parties must accomplish together, under an arrangement that gives them starkly unequal power to do so. Once design is placed beyond the institution's reach, the question stops being which fairness method performs best and becomes what fairness work is even possible, and for whom. The ASP-HEI cycle (Section~\ref{sec:background}) predicts that adopting algorithms under these conditions will deepen existing inequities and concentrate institutional power \cite{mcconvey_this_2024}; we test whether the post-hoc methods the literature recommends for constrained settings can interrupt that pattern or whether they reproduce it.

We approach this coordination problem through the form fairness actually takes for the people doing the work. Advisors never see statistical parity or equalized odds; they see caseloads. A false positive is a struggling student who never lands on any advisor's list. A false negative is a successful student taking up a scarce advising slot. Whatever criterion a fairness intervention optimizes, it ends up moving these two errors around between groups, and with them, which students the institution can see well enough to help. So we judge each intervention not just by whether its fairness metrics converge, but by how it redistributes these errors. This surfaces a pattern we call \emph{fairness theatre}: post-hoc adjustments that make the metrics on an institution's dashboard look fair while the errors students and advisors actually live with stay the same or get worse. Fairness theatre isn't intentional deception; it is what procurement produces when it confines fairness work to the few outputs a vendor exposes and keeps the design decisions that create the disparities out of the institution's reach.

We draw on administrative data covering 168,550 student records from a publicly supported Ontario college. We evaluate the interventions without accessing, modifying, or retraining the underlying predictive model. Five implementations use calibrated probability scores together with demographic-group labels and, where optimization requires them, observed outcomes. Our Bias Mitigation adaptation additionally uses institutional input features to train surrogate models, but it likewise requires no access to the underlying model's parameters, architecture, or training procedure. One mathematical constraint shapes everything that follows: when demographic groups succeed at different base rates, no method can satisfy calibration, equalized odds, and statistical parity at once \cite{kleinberg_inherent_2016, chouldechova_fair_2017} (see Section~\ref{sec:limitations}). The existence of these trade-offs is not new; impossibility results guarantee them, and prior benchmarking has documented the fairness--accuracy tensions and redistribution effects that follow, under conditions where researchers have full access to the model. Our contribution is to show what these familiar dynamics look like once intervention is prohibited from modifying or interrogating the underlying model.

Our research questions follow from this constraint:
\begin{itemize}
    \item \textbf{RQ1: Fairness Trade-offs and Their Distribution.}
    How do different post-hoc methods distribute the mathematically inevitable trade-offs between fairness metrics, and which demographic groups bear the costs of each method's optimization choices?
    \item \textbf{RQ2: Fairness--Accuracy Trade-off Magnitude.}
    What are the magnitudes of fairness--accuracy trade-offs for each method, how do these interact with institutional constraints such as advising capacity, and how do they reshape the error burdens that advisors and students encounter in practice?
    \item \textbf{RQ3: Impacts on Students at the Margins.}
    How do post-hoc interventions affect students at the margins of institutional data systems—particularly those in small or underrepresented demographic categories whose treatment reveals the limits of single-axis fairness frameworks?
\end{itemize}

This study is a stress test, not a prescription. We do not define what a fair EWS should look like or trace disparities to particular causes. We take the post-hoc tools that fairness literature recommends and run them under the constraints procurement actually imposes, evaluating fairness across three established metrics: statistical parity, equal opportunity, and equalized odds (Section~\ref{Data and Modeling}). Because our analysis rests on a single institution and simulated procurement constraints, we read the findings as indicative of how these interventions behave in similar vendor-controlled settings rather than as settled facts about all of them.

The paper makes three contributions to CSCW. First, we reframe post-hoc fairness in procured AI as a problem of cooperative work: fairness is not a property of a model but an outcome negotiated among vendors, institutions, advisors, and students, and procurement contracts restrict the mutual visibility and control this negotiation requires. Second, we introduce error-type profiling, a lens that translates fairness metrics into the terms advisors actually encounter: which students appear on caseloads and which are never seen. Third, we develop the concept of fairness theatre empirically, showing how post-hoc adjustments can satisfy dashboard-level fairness metrics while leaving the distribution of error burdens intact or worse. Two of the six methods we evaluate directed corrections toward already-advantaged groups, a failure whose immediate cause is how our implementations operationalized disadvantage using group size, and which persists because the informational constraints of procurement leave institutions unable to detect or override it.

\begin{figure}
    \centering
    \includegraphics[width=0.75\linewidth]{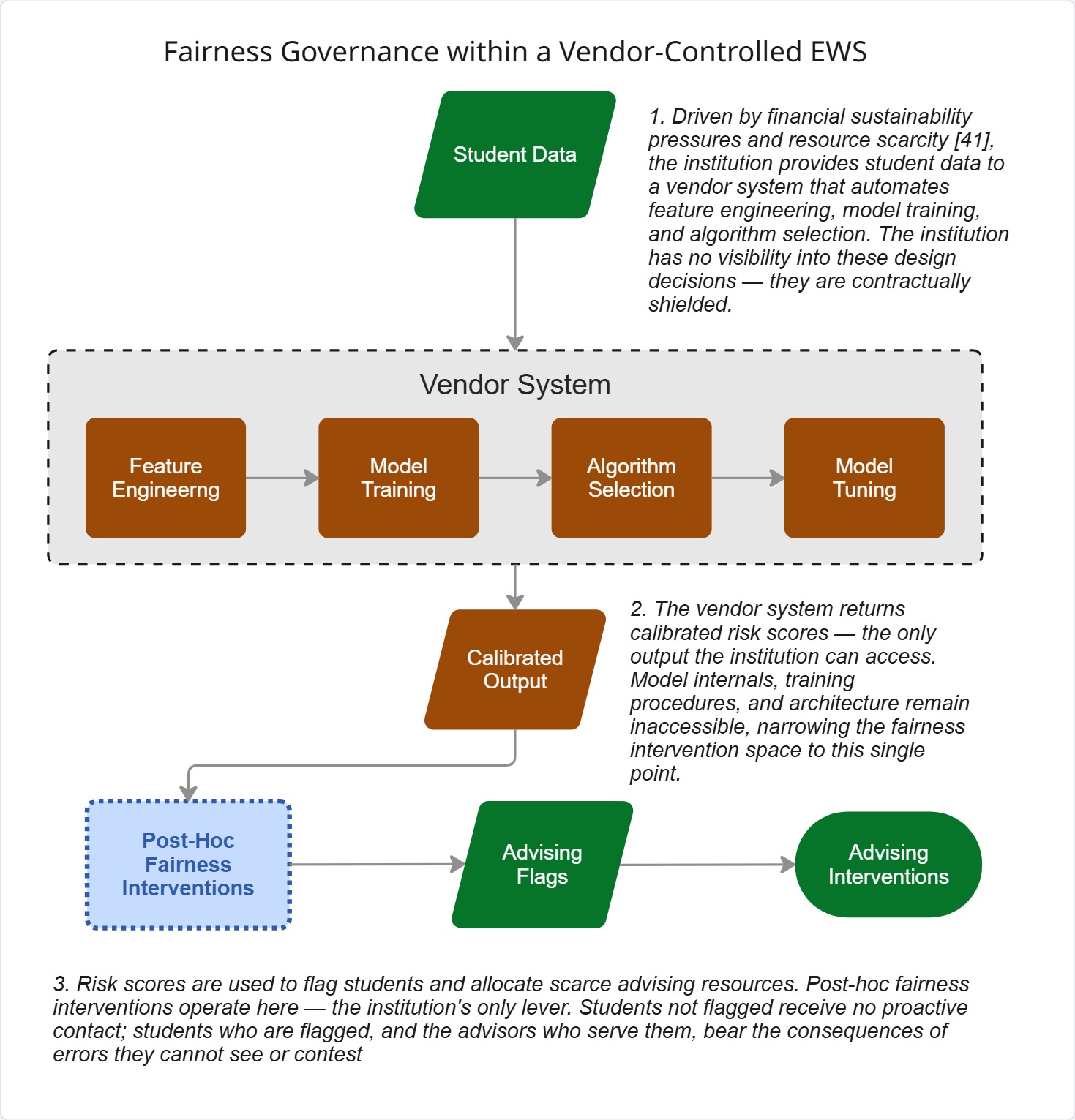}
    \caption{Fairness governance within a vendor-controlled EWS. The vendor controls the predictive model and its design decisions (shaded region); the institution's available levers are limited to post-hoc adjustments made outside the model, primarily through calibrated outputs and, for some interventions, institutional feature data. Advisors act on adjusted flags, and students who are not flagged receive no proactive contact. The contract restricts the flow of knowledge in both directions: vendors do not see the institutional context in which their scores allocate advising work, and institutions cannot interrogate the design decisions that produce the scores. This paper examines what fairness work is possible within the narrow intervention space this arrangement creates.}
    \Description{Fairness governance within a vendor-controlled EWS}
    \label{fig:post_hoc}
\end{figure}

\section{Background: Early Warning Systems Under Procurement}\label{sec:background}
Consider an advisor at a mid-sized public college in the first weeks of the fall term. Her dashboard, populated by a vendor-developed EWS, lists forty students flagged as at risk of not completing their programs. She will spend the next month working through this list: scheduling meetings, drafting support plans, connecting students to tutoring and financial aid counseling. Two students illustrate what the list does and does not show her. The first is a nineteen-year-old domestic student, the first in her family to attend college, flagged by the model largely on the basis of attributes she cannot change: her postal code, her funding source, her first-generation status. She is in fact managing well, but the flag consumes one of the advisor's scarce slots and arrives, from the student's perspective, as an institutional judgment that she does not belong. The second student is an international student in his first semester, struggling quietly with coursework in his third language. The model, trained on historical completion data in which international students completed at high rates, scores him as likely to succeed. He appears on no list. The advisor never learns he exists until he fails to register for the winter term.

Every fairness intervention this paper evaluates redistributes students between these two positions: onto lists and off of them. The interventions differ in which students they move, and in which direction. What they share is the constraint under which they operate: the institution deploying them cannot see why the model scored either student as it did, cannot change how it does so, and cannot ask. This scenario is a composite constructed for illustration. It is grounded in prior ethnographic fieldwork at our research site documenting advisor workflows around EWS flags \cite{mcconvey_this_2024}, and in the group-level patterns we report in Section~\ref{sec:findings}, where domestic and traditional-age students faced the highest flag rates and international students consistently received the highest predicted success rates. It does not describe specific individuals.
\subsection{How EWS Enter and Organize Institutional Work}

In higher education, vendor-controlled AI enters institutional workflows through two pathways: turnkey platforms (e.g., EAB Navigate360, Civitas Learning, among others) that deliver predictions without institutional involvement in model construction, and AutoML platforms (e.g., DataRobot, H2O.ai) that automate algorithm selection and feature engineering \cite{frank_hutter_automated_2019, baratchi_automated_2024}. Both constrain institutional oversight \cite{xin_whither_2021}. The data scientists responsible for building our research site's predictive models describe having limited visibility into the platform's automated processes and no ability to manually adjust the selected algorithms, with model improvement constrained to the addition of new data features \cite{mcconvey_this_2024}. AutoML optimization typically prioritizes accuracy over equity \cite{drozdal_trust_2020, weerts_can_2024}, and procurement constraints prevent auditing whether proxy discrimination operates in vendor-supplied models \cite{barocas_big_2016}. Colleges increasingly rely on both types of systems in implementing EWS to predict student success and allocate scarce advising resources \cite{mcconvey_human-centered_2023}. Post-hoc fairness interventions become relevant precisely because they operate on outputs, the only component vendors consistently expose \cite{chen_post-hoc_2024}.

\subsection{The ASP-HEI Cycle}
These procurement dependencies are not accidental but structurally produced. Prior ethnographic research documented how external policy pressures toward financial sustainability combined with chronic resource scarcity drive higher education institutions (HEIs) into increasing reliance on vendor-supplied algorithmic systems, a dynamic theorized as the ASP-HEI Cycle (Algorithms, Student Data, and Power in Higher Education Institutions) \cite{mcconvey_this_2024}. That cycle describes how the adoption of data-driven practices redistributes institutional power toward business divisions, increases surveillance of students, exacerbates existing inequities, and automates relationships previously mediated by faculty judgment, and how the resulting consolidation of institutional power creates conditions for the cycle to repeat. 

This study examines one concrete governance site within this cycle: the moment when an institution, already locked into a vendor-controlled predictive system, attempts to enact fairness on outputs it can observe but cannot meaningfully interrogate. Our contribution is to show empirically what fairness governance looks like at this site and why the structural constraints the ASP-HEI cycle identifies render post-hoc interventions insufficient. Throughout this paper we refer back to this cycle as the structural context within which the fairness work we evaluate takes place, rather than re-deriving it at each point of use.

\subsection{The Intervention Space}

When institutions procure vendor-supplied models, they typically retain access to two points of intervention: the input data fed into the model, and the calibrated outputs returned by it, probability scores adjusted for reliability. However, vendor contracts generally prohibit access to or modification of the model itself, including its training procedures, internal parameters, and architecture. In this study, we focus on calibrated outputs as the primary fairness lever, for two reasons. First, while input data can be curated or augmented, changes to inputs do not guarantee predictable changes in a model whose internals are inaccessible. Second, calibrated outputs represent the actionable decision point: the classifications that advisors see, that determine which students are flagged, and that drive the allocation of scarce resources. This contractual reality dramatically narrows the intervention space. Most fairness research assumes practitioners can modify training procedures or adjust internal parameters, but vendor procurement explicitly forbids these capabilities. A growing body of work positions post-hoc fairness interventions, adjustments applied to model outputs after deployment without modifying the model itself, as viable strategies for such constrained settings \cite{chen_post-hoc_2024, lohia_bias_2018, sikdar_getfair_2022}. Yet how these methods perform under real procurement constraints, with heterogeneous student populations and operational resource limits, remains largely untested. Figure~\ref{fig:post_hoc} summarizes this arrangement. The shaded region marks decisions the institution cannot see or change: feature selection, outcome definition, training-data composition, optimization targets, and model parameters. The unshaded region contains the resources that remain available outside the vendor model: calibrated probability scores, demographic-group labels, institutional feature data already held by the institution, classification thresholds, and the advising work generated by the resulting flags. Most interventions in our evaluation act directly on calibrated scores. Our Bias Mitigation adaptation also uses institutional feature data to construct a surrogate model, but it does not access, query, modify, or retrain the underlying predictive model.

\subsection{Research Context}
This research was conducted in collaboration with a mid-sized, publicly funded college in Ontario, Canada, one of the province's 24 publicly assisted colleges \cite{government_of_ontario_published_2023, ontario_colleges_library_service_ocls_student_2024}. Over the past decade, public funding for Ontario colleges has declined from 54\% to 32\%, increasing reliance on student fees \cite{government_of_canada_trends_2022}. International enrollment grew six-fold between 2009 and 2019 \cite{government_of_canada_trends_2022} until a federal visa cap in 2024 \cite{crawley_ontarios_2023}. This transformation has driven the adoption of vendor-developed predictive analytics for student retention, raising fairness questions when systems are developed externally with limited institutional oversight.

\section{Related Work}
This section situates our study at the intersection of two bodies of literature that rarely meet: technical work on fairness interventions, which largely assumes full access to models and data, and CSCW scholarship on how cooperative work is coordinated across organizational boundaries, which has examined procurement and algorithmic systems but rarely fairness methods themselves. We first establish the theoretical limits of statistical fairness optimization (\ref{sec:limitations}) and the stakes of predictive systems in higher education (\ref{higher_ed}). We then reframe procurement as a coordination problem for fairness work (\ref{procurement}), and review the post-hoc methods proposed for exactly the constrained settings procurement creates (\ref{constraints}). The gap our study fills sits at the junction: the fairness literature recommends post-hoc methods for constrained settings but has not tested them under procurement conditions, while the CSCW literature has documented procurement's accountability gaps but not what fairness work remains possible within them.

\subsection{Limitations of Statistical Fairness Optimization}\label{sec:limitations}
Extensive scholarship has demonstrated that statistical optimization approaches to algorithmic fairness are insufficient for addressing structural injustice \cite{green_algorithmic_2020,kasirzadeh_algorithmic_2022,zhang_affirmative_2022,hanna_towards_2020}. We highlight three critical limitations. First, fundamental mathematical constraints restrict what fairness interventions can achieve. \citet{kleinberg_inherent_2016} proved that when groups have different base rates, it is impossible to simultaneously satisfy calibration, equalized odds, and statistical parity; \citet{chouldechova_fair_2017} demonstrated a parallel impossibility for predictive parity and equal error rates. These results mean no post-hoc method can achieve perfect parity across all metrics we evaluate. Second, statistical fairness metrics focus on distributing outcomes across groups but fail to account for \textit{structural injustice}. \citet{kasirzadeh_algorithmic_2022} argues that even a ``fair'' distribution may have negligible social significance when underlying power dynamics go unaddressed. \citet{zhang_affirmative_2022} extends this through relational equality, showing that equal error rates do not ensure equal justice when the basic structure is unjust. Third, fairness methodologies systematically treat protected attributes as ``fixed'' rather than ``structural, institutional, and relational'' \cite{hanna_towards_2020}, minimizing structural aspects of unfairness. Simply disaggregating metrics by demographic categories does not address how data collection practices and outcome definitions encode stratification.

A growing body of scholarship reinforces these critiques. \citet{selbst_fairness_2019} show that fairness interventions often fall into an ``abstraction trap,'' optimizing technical properties divorced from social context, and \citet{jacobs_measurement_2021} show that choosing which outcomes to predict and how to define success encodes contested normative assumptions rather than neutral ground truth \cite{bao_its_2022, wang_aleatoric_2023, weerts_neutrality_2024, vethman_fairness_2025}. \citet{green_algorithmic_2020} name this common thread as algorithmic formalism's ``narrowing of vision'': the political choice of which parts of justice a metric captures and which fall outside the frame.

Our work sits within this tradition. We do not claim fairness metrics reveal structural injustice but offer what \citet{bao_its_2022} call a ``critical empirical study'', a stress test of the post-hoc methods this literature recommends for constrained settings, run under the conditions institutions actually face. Where the work above assumes full access to models and data, we restrict intervention to vendor-controlled outputs. These theoretical limits shape what any statistical fairness intervention can achieve. In the higher education context we examine, procurement relationships and AutoML platforms compound them, as we discuss next.

\subsection{AI in Higher Education}\label{higher_ed}
HEIs collect extensive data on students, from sociodemographic profiles and financial records to academic performance histories \cite{mcconvey_human-centered_2023}. Advances in computing power and machine learning have enabled these data sources to be used for decision-making at scale. AI applications in this domain range include learning analytics dashboards \cite{sabuncuoglu_developing_2023}, remote exam proctoring \cite{kaliisa_have_2024}, and outcome prediction \cite{hermogenes_analysis_2024}. Critical scholarship has emphasized the ethical stakes of this shift, raising concerns about data governance, privacy, surveillance, and the institutional responsibility to ensure fairness-aware design \cite{slade_learning_2013,prinsloo_elephant_2017}. Fairness is not merely a technical measure in this setting; it becomes a policy mechanism for embedding social values into decision-making processes \cite{abebe_roles_2020}.

EWS, introduced in Section~\ref{sec:background}, are among the most consequential of these applications because their predictions allocate scarce support resources. Their development is often constrained by departmental data silos, limited resources, and uneven access to data science expertise \cite{mcconvey_this_2024}. \citet{perdomo_difficult_2023} show that including sensitive attributes does not necessarily improve predictive performance: community-based models in Wisconsin public schools matched individual-level models while avoiding explicit use of race and gender. Meanwhile, biases in training data, including class imbalances and correlations with protected attributes, can become codified as ground truth. These dynamics can produce what \citet{andalibi_conceptualizing_2023} term \textit{algorithmic stigmatization}: incorrect predictions trigger unnecessary interventions that cause stress and stigma, while missed identifications deprive students of needed support. Yet most institutions do not build these systems in-house, which raises the question of what happens to fairness when the system arrives from outside.

\subsection{Procurement as a Coordination Problem for Fairness Work} \label{procurement}
Across public sector contexts, there has been a marked shift from internally developed AI systems to vendor-supplied solutions. Research from HCI, policy institutions, and case studies surface the same problems when institutions adopt commercial ``black box'' systems: the systems are not transparent, no one can be held accountable for them, and they carry systematic bias \cite{kapania_im_2025,bartosiak_fired_2022,casper_black-box_2024,abdul_trends_2018}. AI tools in public institutions are now acquired through procurement rather than in-house development, with procurement frameworks originally designed for goods and basic services now structuring decisions about algorithmic governance \cite{johnson_public_2024}.

This shift is not limited to formal contracts. Alternative procurement pathways such as low-cost purchases that bypass oversight, subscription models, and philanthropic donations enable the quiet introduction of AI features into existing systems without triggering review \cite{sattlegger_beyond_2024,wagenknecht_algorithms_2016,xu_toward_2019}. Faced with shortages in technical expertise, many public agencies view vendor procurement as the only viable option \cite{veale_fairness_2018}. Yet this reliance deepens information asymmetries: purchasers often lack the capacity to evaluate systems or negotiate effectively with large technology firms, while vendors invoke trade secret protections to shield algorithms, training data, or decision logic from scrutiny \cite{selbst_disparate_2017,citron_technological_2008}. Even when audit rights exist on paper, technical complexity and restricted access frequently render meaningful oversight infeasible \cite{raji_closing_2020,metcalf_algorithmic_2021}. The result is what \citet{pasquale_black_2015} calls “black box governance,” where substantive policy decisions are effectively made by private companies through design choices, often without democratic input \cite{mulligan_procurement_2019}. 

CSCW offers conceptual resources for understanding what procurement does to fairness. A long tradition in the field treats cooperative work as depending on \emph{articulation work}: the constant effort of keeping distributed tasks, people, and understandings aligned \cite{schmidt_taking_1992}. Procurement contracts cut exactly the channels that alignment needs. The vendor holds knowledge of features, training data, and optimization targets; the institution holds knowledge of student populations, advising capacity, and equity obligations; the contract ensures neither party's knowledge reaches the other's decisions. Scholarship on infrastructure shows how embedded, taken-for-granted systems structure what work is possible long before any individual task begins \cite{star_steps_1996}, and invisible work demonstrates how the labor and the people that systems do not represent tend to disappear from organizational view \cite{star_layers_1999}. Both dynamics recur in our findings: the procurement arrangement preconfigures the intervention space, and the students least represented in institutional data are the ones fairness methods serve worst.

CSCW and HCI literature examines fairness as situated organizational practice rather than model property. \citet{holstein_improving_2019} document that industry practitioners' fairness needs diverge sharply from what the technical literature provides, including limited access to the data and models they are asked to make fair. \citet{veale_fairness_2018} find that public sector machine learning practitioners work under institutional constraints that fairness research rarely models. Studies of algorithmic decision support in social services show how frontline workers absorb, contest, and repair the outputs of systems they did not design \cite{kawakami_improving_2022, saxena_algorithmic_2024}. Our study extends this line of work in two ways: we examine fairness work under contractual rather than merely practical constraint, and we evaluate what the recommended technical remedies actually deliver from that position.

\subsection{The Constrained Fairness Intervention Landscape}\label{constraints}

If procurement severs the coordination that fairness work requires, what tools remain to an institution that still bears responsibility for equitable outcomes? Fairness literature's answer is post-hoc intervention, and a body of benchmarking research has tested such methods under controlled conditions \cite{l_cardoso_framework_2019, reddy_benchmarking_2021}. That work is valuable but consistent in its blind spot; it assumes full access to training data and model internals and it finds that even state-of-the-art methods degrade when sensitive attributes correlate with outcomes or when marginalized groups are under-represented. Procurement removes exactly the access these benchmarks presuppose, further complicating the fairness problem.

Beyond technical interventions, institutions might address procurement constraints through contractual mechanisms such as negotiating stronger audit rights, requiring pre-deployment bias testing, or developing systems in-house. But these alternatives run into compounding barriers. Many institutions cannot evaluate a vendor's fairness claims or write meaningful algorithmic requirements into a contract in the first place \cite{veale_fairness_2018}. The concentrated educational technology market gives individual institutions little leverage to demand stronger oversight terms \cite{mulligan_procurement_2019}, and resource-constrained public institutions struggle to sustain the data science expertise required for in-house development \cite{mcconvey_this_2024}. Even contractually guaranteed audit rights often prove ineffective as technical complexity and limited vendor cooperation render formal access rights practically meaningless \cite{raji_closing_2020,metcalf_algorithmic_2021}. We focus on post-hoc interventions because they represent the intervention space structurally available across both procurement pathways, and because a growing body of fairness research recommends them for deployment-constrained settings \cite{chen_post-hoc_2024, lohia_bias_2018, sikdar_getfair_2022, angelopoulos_gentle_2022}. Our study evaluates whether these recommendations hold under realistic conditions: heterogeneous student populations, unequal base rates, small and underrepresented categories, and operational resource limits.

\section{Methods}
To systematically evaluate post-hoc fairness interventions under procurement constraints, we developed a comparative experimental framework using institutional data from a large Canadian college, implementing six distinct fairness methods on calibrated outputs from a replicated EWS.

\begin{table*}[t]
\centering
\footnotesize
\caption{Summary of post-hoc fairness methods evaluated. All methods adjust model outputs without retraining or accessing the model's internal structure.}
\label{tab:fairness-summary}
\begin{tabular}{
    p{0.12\textwidth}
    p{0.20\textwidth}
    p{0.18\textwidth}
    p{0.17\textwidth}
    p{0.20\textwidth}
}
\toprule
\textbf{Method} & \textbf{Approach} & \textbf{Primary Optimization Target} & \textbf{Source} & \textbf{Key Limitations} \\
\midrule

\hyperref[getfair]{GetFair} &
Learns a single global decision threshold via reinforcement learning-inspired optimization &
Configurable: SP, EOp, or EO (via reward function) &
Sikdar et al.\ (2022) \cite{sikdar_getfair_2022} &
Adjusts one attribute at a time; extreme classifications for small groups; substantial accuracy loss at high fairness weights \\
\addlinespace

\hyperref[decoupled]{Decoupled classifiers} &
Trains group-specific logistic regression models on base model probability scores (post-hoc adaptation) &
Configurable (joint accuracy--fairness loss) &
Dwork et al.\ (2018) \cite{dwork_decoupled_2018} &
Requires large per-group samples; reverts toward pooled baseline for small groups; cannot handle intersections \\
\addlinespace

\hyperref[exploiting]{Reject option fairness} &
Relabels uncertain predictions near the decision boundary based on group membership &
Statistical parity &
Kamiran et al.\ (2018) \cite{kamiran_exploiting_2018} &
Effectiveness depends on density of predictions near boundary; minimal adjustment for groups with few borderline cases \\
\addlinespace

\hyperref[bias scoring]{MBS} &
Learns bias scores from group-conditional models; selectively flips predictions exceeding bias threshold &
Equalized odds (with accuracy floor) &
Chen et al.\ (2024) \cite{chen_post-hoc_2024} &
Requires group labels and observed outcomes during configuration; results depend on the selected fairness constraint and accuracy floor\\
\addlinespace

\hyperref[bias mitigation]{Bias mitigation post-processing} &
Audits predictions for group-level disparities and selectively corrects largest imbalances &
SP, EOp, and EO simultaneously &
Lohia et al.\ (2018) \cite{lohia_bias_2018} &
Our adaptation requires institutional input features and observed outcomes to train surrogate models; surrogate predictions may not reproduce the inaccessible model's counterfactual behaviour \\
\addlinespace

\hyperref[conformal]{GBCP} &
Adapts group-balanced conformal prediction to set group-specific probability thresholds that equalize positive prediction rates to a shared target &
Statistical parity (adapted from coverage equalization in original) &
Angelopoulos \& Bates (2022) \cite{angelopoulos_gentle_2022} &
Original method equalizes coverage (prediction set reliability) across groups; our adaptation repurposes the quantile-threshold mechanism to target statistical parity, which produces uniform positive rates regardless of base rates; we use 0.50 target rate \\
\bottomrule
\end{tabular}
\end{table*}

\subsection{Study Design and Its Limits}\label{sec:observed_simulated}
Prior ethnographic fieldwork at our research site \cite{mcconvey_this_2024} documented the operational EWS: the institution procured an AutoML platform that selected tree-based ensemble models, staff could see little of its automated pre-processing and could not change the algorithms it chose, and advisors received risk assessments that shaped their caseloads. The data we analyze is the institution's real student data, and the outcome we predict is its own administrative definition of success.

The interventions themselves could not be run on the operational vendor model, so we simulated the institution's position. We built a research replica, an XGBoost classifier of the same model class the AutoML platform had selected, trained on the same data with SMOTE-NC class balancing and post-hoc calibration. During the intervention stage, we treated the replica as inaccessible: none of the six methods could inspect its parameters or architecture, alter its training procedure, or retrain it. Most methods operated on calibrated probability scores, demographic-group labels, and observed outcomes used during configuration. The Bias Mitigation adaptation additionally used institutional input features to train surrogate models outside the replica.
We cannot verify that the replica's individual predictions match the operational system's, and we do not claim they do. What it preserves is the structural position of the institutional user, acting on calibrated scores from a tree-ensemble model trained on this population without the ability to see or change how they were produced.

Anything we say about vendor-controlled systems in general is an inference from this setup, and we are careful about how far it extends. Findings that turn only on the structural position should carry over to other procurement settings: the narrow room to intervene, the way methods must guess at who is disadvantaged, the impossibility of certain fairness constraints once base rates differ. Findings that turn on our replica's particular error distribution, like the exact TPR and FPR values, we do not expect to carry over, and we offer them as a sense of scale rather than estimates of any real vendor system. Our replica is a favourable test case; it had the benefit of SMOTE-NC balancing that a vendor pipeline may lack, so methods that struggle here may struggle more against models trained on imbalanced data.

Prior fieldwork at the college provides qualitative grounding for the simulated setting: staff described an AutoML platform that selected tree-based models, limited visibility into model development, and an Early Alert System whose classifications shaped advising work \cite{mcconvey_this_2024}. That study did not report production flag rates, score distributions, calibration statistics, or subgroup prediction patterns, and we had no such aggregates available for comparison. We therefore cannot assess whether the replica resembles the operational system's prediction distribution. The fieldwork supports the plausibility of the workflow and institutional constraints represented here, rather than equivalence between the two models' outputs.

\subsection{Algorithm Development} \label{Data and Modeling}
Following approval by the Research Ethics Boards at both partnering institutions (anonymized for review), we obtained 15 years of institutional student data, comprising demographic, admissions, and registration records. The source dataset, provided by the institution, included 875,697 student-semester records spanning from 2008 to 2023 and covering semesters 1 through 6 of students' programs. All personally identifiable information was removed by the institution prior to data access. Student IDs were anonymized via hashing to ensure privacy and longitudinal linkage.

To simulate a realistic EWS use case, we restricted the dataset to students entering their first academic semester, retaining only features that would be available at the time of intake (e.g., excluding end-of-term academic outcomes but retaining High School GPA). This intake-only design reflects the institutional context at our research site, where predictive risk scores generated at enrollment are intended to inform resource allocation decisions \cite{mcconvey_this_2024}. While operational EWS often update predictions throughout the academic term using mid-semester grades and attendance data, our focus on intake prediction serves two purposes: (1) it captures the highest-stakes moment when institutions stratify students into support tiers before academic performance data exists, and (2) it mirrors the procurement scenario where vendors often deliver risk scores at enrollment based on pre-matriculation data, with mid-term updates requiring additional licensing or manual re-scoring. This design choice prioritizes evaluating fairness at the point of initial triage rather than ongoing monitoring. After restricting the data to first-semester intake records and excluding records whose target outcome was missing or coded as ``ZZ - Unknown,'' the modeling cohort contained 168,550 student records: 97,599 Domestic students and 70,951 International students.

\subsubsection{Data Pre-processing}
Categorical variables were encoded using appropriate methods (e.g., ordinal encoding or one-hot encoding, where categorical variables are converted into binary indicators). We removed columns that were either irrelevant to prediction tasks or likely to introduce data leakage (e.g., post-intake academic outcomes). The final set of features included:

\begin{itemize}
    \item \textbf{Target variable:} binary outcome indicating academic success (completed their program within its allowable period) or non-success (withdrawal, academic failure, or transfer). This is the college's existing administrative definition, used for enrollment management and resource planning, and we adopt it to preserve ecological validity. It is also a consequential simplification: it conflates financially driven withdrawal, program mismatch, transfer to another institution, and academic failure into a single ``unsuccessful'' label, treating structurally produced outcomes as individual risk \cite{baker_algorithmic_2022}. Ground truth here is thus an institutional artifact, not a neutral fact about students, and every fairness metric we report inherits this definition. A further caveat: the college has historically operated early warning tools and support programs, but we lack records of which students received interventions, so some outcome labels reflect intervention effects we cannot model, a common limitation in retrospective EWS analyses. Because the task is to flag students at elevated risk, we evaluate ROC-AUC using inverted probability scores ($1 - P(\text{success})$) so that the positive class corresponds to at-risk students.
    \item \textbf{Program features:} program code, credential type, program length, campus, and program cluster.
    \item \textbf{Admissions features:} High School GPA, prior postsecondary education and GPA, English reading and speaking scores.
    \item \textbf{Residency features:} citizenship status, permanent country of residence, immigration landed date, first language, and partial postal code.
    \item \textbf{Sensitive features:} age, gender, funding source, and first-generation status. These are the demographic attributes available in the institutional dataset; age and gender are standard fairness-audit categories, while funding source and first-generation status are commonly used equity indicators in higher education. These attributes are included as predictive features, replicating the design of the operational EWS at our research site. We include them for three reasons. First, because our study evaluates fairness interventions under simulated procurement conditions, the model must reflect the feature set institutions actually encounter; omitting attributes the institutional model uses would undermine ecological validity. Second, excluding protected attributes does not prevent proxy discrimination: variables such as postal code, first language, and funding source can encode demographic information even when protected attributes are formally absent \cite{perdomo_difficult_2023, barocas_big_2016}. Explicit inclusion can support direct measurement of group-level disparities where the attribute data are sufficiently complete rather than allowing them to operate through unmeasured proxies. Third, Canadian human rights frameworks and institutional equity policies permit demographic data use for equity-promoting purposes, the legal concern being discriminatory \textit{outcomes} rather than data use per se. Using protected attributes in an opaque model nonetheless raises legitimate concerns, which apply to the institution's design choice that our study replicates rather than endorses.

\end{itemize}
    
In total, the processed dataset included 46 predictive features.

\subsubsection{Handling of Demographic Categories}

We selected age, gender, and residency status as the three axes for the fairness analysis because the institution provided sufficiently complete and usable data for these attributes to support group-level comparisons. Funding source and first-generation status were also available as predictive features, but substantial data-quality problems made their categories unsuitable for the same analysis (Funding Source was not meaningfully categorized beyond residency status; First Generation Status was reported for only 47\% of students). Their exclusion limits the scope of our findings: disparities along these attributes, and at their intersections with the three analyzed axes, remain unassessed.

The following counts describe the full 168,550-record modeling cohort before partitioning and before the application of SMOTE-NC:
\begin{itemize}
    \item \textbf{Gender:} Female ($n=\textnormal{83{,}660}$), Male ($n=\textnormal{84{,}736}$), Unknown Gender ($n=\textnormal{154}$)
    \item \textbf{Age Group:} Under 25 ($n=\textnormal{116{,}775}$), 25--35 ($n=\textnormal{32{,}416}$), Over 35 ($n=\textnormal{16{,}259}$), Unknown Age ($n=\textnormal{3{,}100}$)
    \item \textbf{Residency:} Domestic ($n=97{,}599$), International ($n=70{,}951$)
\end{itemize}

Age was collapsed from continuous values into three categories reflecting institutional classifications of student populations: ``Under 25'' captures traditional-age students typically entering directly from secondary or other post-secondary education, ``25-35'' represents early-career and returning students, and ``Over 35'' captures mature students often balancing education with caregiving or employment responsibilities. Records without usable age information were retained as a separate ``Unknown Age'' category ($n=3{,}100$ in the full modeling cohort). We included this category in the age-based fairness analyses rather than dropping these records.

Within the data provided to us, students with non-binary gender identities, those who selected ``Other'' or ``Prefer not to say,'' and those with missing gender data were consolidated into a single ``Unknown Gender'' category (total $n=154$). We acknowledge this aggregation obscures important differences in experiences across non-binary, genderqueer, agender, and other gender-diverse identities, as well as conflating students who actively declined to report gender with those for whom data was not collected. This limitation, discussed further in Section \ref{sec:limitations-directions}, reflects how institutional data collection practices can render marginalized students statistically invisible. The unknown gender category's small sample size (total $n=154$, representing 0.09\% of the modeling cohort) has important implications for fairness evaluation as metrics for this group carry far more uncertainty than for the binary gender categories (see \ref{sec:comparison}). 

\paragraph{Evaluation Strategy by Method Type.}
Methods handled small groups differently: those requiring group-specific modeling (Decoupled, GBCP) produced unstable estimates for unknown gender; threshold-based methods (GetFair, Reject Option) computed separate thresholds for all groups; score-based methods (MBS, Bias Mitigation) adjusted predictions but lacked power for small groups. We report all results as produced, while flagging that metrics for very small groups ($n < 100$) warrant caution.

\subsubsection{Modeling and Evaluation}
We partitioned Domestic and International records separately using a fixed random seed. Within each cohort, approximately 50\% of records were assigned to the training partition, and the remaining records were divided equally between calibration and test partitions. This produced 84,274 observed training records, 42,138 calibration records, and 42,138 held-out test records. Table~\ref{tab:data-partitions} provides the exact counts by cohort.

Because the training data contained unequal numbers of successful and unsuccessful students, we applied SMOTE-NC only to the training partition. SMOTE-NC created synthetic examples of the minority class while preserving the structure of categorical variables such as program type and citizenship status. Neither the calibration partition nor the test partition was resampled.

\begin{table}[t]
\centering
\small
\caption{Observed records in the modeling cohort and each data partition before the application of SMOTE-NC. The calibration and test partitions contain no synthetic observations.}
\label{tab:data-partitions}
\begin{tabular}{@{}lrrrr@{}}
\toprule
\textbf{Cohort} &
\textbf{Total} &
\textbf{Training} &
\textbf{Calibration} &
\textbf{Test} \\
\midrule
Domestic      & 97,599  & 48,799 & 24,400 & 24,400 \\
International & 70,951  & 35,475 & 17,738 & 17,738 \\
\midrule
Total         & 168,550 & 84,274 & 42,138 & 42,138 \\
\bottomrule
\end{tabular}
\end{table}

We employed an XGBoost classifier for predictive modeling. Prior ethnographic research at this research site documented that the institution's AutoML platform selected tree-based ensemble models with limited visibility into the specific algorithm chosen \cite{mcconvey_this_2024}; our use of XGBoost replicates this model class. Separate models were trained and calibrated for Domestic and International student subsets as the admissions processes and consequently, the available features, differ substantially between the two groups. This approach also aligns with the research site's current practices, which treat Domestic and International cohorts as distinct populations for analysis. The XGBoost models were fitted on the training partition. Separate isotonic probability calibrators for the Domestic and International models were fitted on the calibration partition. Where a post-hoc method required parameter estimation, threshold selection, group-specific model fitting, or another data-dependent configuration, that configuration was determined using the calibration partition and then fixed before evaluation. We applied the calibrated models and fixed post-hoc interventions to the untouched test partition. All final accuracy, fairness, and error-redistribution results reported in Section~\ref{sec:findings} were computed on the 42,138-record test partition.

\paragraph{Fairness Metrics in the EWS Context}
Unless otherwise stated, all results in this section were computed on the held-out test partition ($n=42{,}138$), comprising 24,400 Domestic and 17,738 International student records. The test partition was not used to fit the XGBoost models, calibrate their probabilities, or select post-hoc intervention parameters. We evaluate three group fairness metrics that reflect different equity concerns in student support allocation:

\textbf{Statistical Parity (Demographic Parity):} Do students from different groups have equal probability of being predicted ``successful''? In EWS, this means asking, for example: Are international and domestic students equally likely to be flagged for support, regardless of their actual outcomes? Large disparities suggest the system may over-surveil certain groups.

\textbf{Equal Opportunity:} Among students who actually succeed, are different groups equally likely to be correctly predicted as successful? This measures the \textit{true positive rate} (TPR)—the proportion of successful students correctly identified. In EWS, unequal TPR means some groups of successful students are more often mislabeled as ``at risk,'' potentially facing unnecessary interventions.

\textbf{Equalized Odds:} This extends Equal Opportunity by also requiring equal \textit{false positive rates} (FPR)—the proportion of unsuccessful students incorrectly predicted as successful. In EWS, high FPR for a group means struggling students are missed and denied support they need. See Section~\ref{fairness metrics} in \autoref{appendix} for formal definitions.

\paragraph{Error Types and the Positive Class.}
Throughout this paper the positive class is \emph{predicted successful}, matching the model's output and our tables; readers accustomed to ``positive = flagged'' should transpose. The two error types carry asymmetric, workflow-specific consequences. A \emph{false negative}, a successful student predicted at-risk, takes a finite advising slot that could have reached a student who genuinely needed it, while risking stigmatization of the flagged student \cite{andalibi_conceptualizing_2023}. A \emph{false positive}, a struggling student predicted successful, is more insidious as the student never appears on the advisor's caseload. We therefore evaluate each method by \emph{how it redistributes errors across types and groups}: whether fairness gains reflect genuine improvement or merely shift which students bear which errors, and which students advisors see. The single exception is ROC-AUC, which, as noted in Section~\ref{Data and Modeling}, we compute on inverted probability scores so that discrimination is reported with respect to the at-risk class; all classification metrics (TPR, FPR, positive rate) follow the predicted-successful convention.

\subsection{Post-Hoc Techniques}
Before applying fairness adjustments, we calibrated the model—a process that ensures the probability estimates are trustworthy. Calibration serves two purposes in our evaluation framework. First, it provides the reliable probability estimates necessary for threshold-based fairness interventions. Second, because we calibrated models separately for Domestic and International student populations (reflecting distinct admissions processes and feature availability), this group-specific calibration can reduce between-group disparities compared to a single global calibration \cite{pleiss_fairness_2017}. However, calibration alone cannot eliminate fairness gaps—particularly within demographic categories like age and gender where no group-specific modeling was performed—which is why post-hoc interventions remain necessary.

All data transformations were implemented in \texttt{pandas}, and auxiliary models were trained using \texttt{scikit-learn}. The six post-hoc fairness methods were implemented as custom Python functions developed from the procedures described in the cited papers (see Table~\ref{tab:fairness-summary}). None requires access to the underlying predictive model's parameters, architecture, or training procedure, and none retrains that model. GetFair, Decoupled Classifiers, Reject Option, MBS, and GBCP operate on calibrated probability scores together with demographic-group labels and, where required for configuration, observed outcomes. Our Bias Mitigation adaptation additionally uses institutional input features and observed outcomes to train surrogate models outside the inaccessible predictive model. See Section~\ref{post-hoc descriptions} in \autoref{appendix} for full implementation details. All six methods operate on a single demographic attribute at a time; this single-axis framework cannot capture compounding disadvantages for intersectional students, a limitation we return to in Section~\ref{sec:representation_paradox}.

\paragraph{Method Selection.}\label{sec:method_selection}
We selected candidate methods by three criteria applied in order. First, \emph{feasibility without model access}: each method had to operate without inspecting, modifying, or retraining the underlying predictive model. Five implementations operate on calibrated scores and associated group information. Bias Mitigation is a partial exception because our adaptation also requires institutional feature data to construct a surrogate model, although it still requires no access to the underlying predictive model. Second, \emph{coverage of fairness criteria}: together the methods target statistical parity, equal opportunity, and equalized odds. Third, \emph{technical diversity}: the set spans threshold tuning (GetFair), group-specific calibration (Decoupled), boundary relabeling (Reject Option), prediction flipping (MBS), audit-and-correct pipelines (Bias Mitigation), and quantile thresholding (GBCP), so that shared failure patterns cannot be attributed to a shared mechanism. We did not select for expected performance.

We initially selected seven methods. One, xOrder \cite{cui_towards_2021}, proved inapplicable on examination, and we report the reasons because they are themselves informative about the distance between fairness methods as published and fairness methods as deployable. First, xOrder is formally defined only for binary sensitive attributes: its cross-group ordering, dynamic program, and fairness bound are all stated for exactly two groups, and every sensitive attribute in the original evaluation is binary. Two of our three attributes fall outside this scope (age has four categories in our data, gender three), and the method specifies no procedure for aggregating across pairwise comparisons into a single per-student prediction. Second, even for our one binary attribute (residency), the published dynamic program is computationally intractable at institutional scale: benchmarking it on our data yields runtime scaling of approximately $n^{2.36}$, which extrapolates to months of computation and over one hundred gigabytes of memory for a single domestic--international comparison at full cohort size. Any faithful application would require either an aggregation rule we would have to invent or a reformulation of the algorithm itself, and either would mean evaluating something other than the published method. We therefore evaluate six methods and treat xOrder's inapplicability as a finding rather than an omission; Appendix~\ref{towards} gives the full benchmarking detail. Table~\ref{tab:fairness-summary} summarizes the six methods evaluated; we describe each below.

\paragraph{GetFair (Generalized Fairness Tuning)} \cite{sikdar_getfair_2022} learns a single global decision threshold optimizing a $\lambda$-weighted combination of fairness and accuracy. Because it applies one threshold across all groups, it can produce extreme classifications for small categories, and at high fairness weights ($\lambda \geq 0.7$) it degrades accuracy substantially.

\paragraph{Decoupled Classifiers} \cite{dwork_decoupled_2018} In our post-hoc adaptation, we train group-specific logistic regressions on the base model's probability scores, treating them as a calibration layer rather than replacing the base model; a transfer-learning parameter $\theta$ borrows strength from larger groups for underrepresented ones. The method requires sufficient per-group samples to learn distinct calibration functions; for very small groups it reverts toward the pooled baseline.

\paragraph{Reject Option Fairness} \cite{kamiran_exploiting_2018} relabels predictions in an uncertainty region near the decision boundary by group membership, favoring disadvantaged groups, to target statistical parity. Its effect depends on how many predictions fall in that region; groups sparse near the boundary receive minimal adjustment.

\paragraph{Modification with Bias Scores (MBS)} \cite{chen_post-hoc_2024} estimates a bias score for each prediction by combining the calibrated probability with an estimate of group membership derived from that probability. It selectively flips predictions whose bias scores exceed a learned threshold, optimizing a fairness criterion subject to an accuracy floor. Our implementation therefore uses calibrated probability scores, demographic-group labels, and observed outcomes during threshold selection, but does not use the full institutional feature set or access the underlying predictive model.

\paragraph{Bias Mitigation Post-processing} \cite{lohia_bias_2018} The original method audits predictions through counterfactual queries to the underlying model. Because such queries are unavailable under our simulated procurement constraints, our adaptation trains surrogate logistic regression models using institutional input features and observed outcomes. It uses these surrogates to approximate counterfactual predictions and selectively replace predictions identified as biased. This adaptation requires more institutional data than the other five implementations but does not query, inspect, modify, or retrain the underlying predictive model. All stochastic components are seeded; we verified seed stability across 20 random seeds, with group-level metrics invariant for gender and residency and varying only in the third decimal for the smallest age category (Appendix~\ref{app:seed_variation}).

\paragraph{Group-Balanced Conformal Prediction (GBCP)} \cite{angelopoulos_gentle_2022} In its original form, GBCP sets group-specific thresholds to equalize coverage, a calibration guarantee rather than a fairness intervention. We adapt the threshold mechanism to a different objective: setting group-specific quantile thresholds so that positive rates converge to a shared target (0.50), directly targeting statistical parity. Because it forces all groups to the same positive rate regardless of base rates, it necessarily degrades TPR for groups whose baseline rates diverge most from the target.

\subsubsection{Operational Realism of Thresholds}\label{sec:realism_thresholds} The calibrated decision thresholds produce baseline flag rates of approximately 8–25\%, depending on cohort and subgroup. This range aligns with typical advising capacity, where only a limited portion of the student body can realistically receive proactive outreach \cite{mcconvey_this_2024}. We therefore treat these thresholds as plausible institutional operating points, and interpret post-hoc fairness results in terms of how they re-allocate limited intervention capacity across student groups. Fairness gains that require substantially increasing the number of flagged students would not be operationally feasible under current resource constraints.

\section{Findings}\label{sec:findings}
Our empirical evaluation reveals differential effectiveness across six post-hoc fairness interventions applied to calibrated EWS outputs. This section presents a systematic comparison of fairness-accuracy trade-offs, first establishing baseline disparities in the calibrated model, then examining how each post-hoc method reshapes group-level outcomes across age, gender, and residency status. We keep this section strictly descriptive—documenting what changed, by how much, and in which direction—reserving normative interpretation for the Discussion. Throughout the following analysis, we interpret results against the operational context described in Section \ref{sec:realism_thresholds}: flag rates must fall within the 8--25\% range that institutional advising capacity can absorb, and fairness gains that require substantially expanding the flagged population represent theoretical improvements without practical viability.
\begin{table*}[ht]
\centering
\caption{Age group fairness metrics across all methods. TPR: True Positive Rate, FPR: False Positive Rate, Pos.\ Rate: Positive classification rate. GetFair reports $\lambda = 0.3$. Bias Mitigation results for the Unknown age group are reported in Section~\ref{sec:bias-mitigation} and Appendix~\ref{app:seed_variation}.}
\label{tab:age-all-methods}
\begin{tabular}{@{}lrrr|rrr|rrr@{}}
\toprule
& \multicolumn{3}{c}{\textbf{Under 25}} 
& \multicolumn{3}{c}{\textbf{25 to 35}} 
& \multicolumn{3}{c}{\textbf{Over 35}} \\
\cmidrule(lr){2-4} \cmidrule(lr){5-7} \cmidrule(lr){8-10}
\textbf{Method} & TPR & FPR & Pos.R 
    & TPR & FPR & Pos.R 
    & TPR & FPR & Pos.R \\
\midrule
Base        
    & 0.883 & 0.339 & 0.721 
    & 0.953 & 0.543 & 0.880 
    & 0.945 & 0.619 & 0.886 \\
Calibrated  
    & 0.938 & 0.430 & 0.787 
    & 0.979 & 0.621 & 0.915 
    & 0.966 & 0.668 & 0.912 \\
\midrule
GetFair     
    & 0.958 & 0.490 & 0.819 
    & 0.988 & 0.673 & 0.932 
    & 0.979 & 0.710 & 0.930 \\
Decoupled   
    & 0.938 & 0.429 & 0.786 
    & 0.979 & 0.621 & 0.915 
    & 0.968 & 0.681 & 0.916 \\
Reject Option 
    & 0.899 & 0.360 & 0.739 
    & 0.963 & 0.559 & 0.892 
    & 0.950 & 0.630 & 0.892 \\
MBS         
    & 0.938 & 0.429 & 0.786 
    & 0.979 & 0.621 & 0.915 
    & 0.966 & 0.668 & 0.912 \\
Bias Mitigation 
    & 0.938 & 0.429 & 0.786 
    & 0.979 & 0.621 & 0.915 
    & 0.966 & 0.668 & 0.912 \\
GBCP        
    & \dgr{0.631} & 0.121 & 0.480 
    & \dgr{0.578} & 0.125 & 0.497 
    & \dgr{0.563} & 0.155 & 0.489 \\
\bottomrule
\end{tabular}

\smallskip
\noindent{\footnotesize 
\colorbox{green!15}{\strut\hspace{0.5em}} Improvement from calibrated baseline ($\Delta$TPR${>}$0.03 or $\Delta$FPR${>}$0.05) \quad \colorbox{red!15}{\strut\hspace{0.5em}} Degradation from calibrated baseline \quad \colorbox{orange!20}{\strut\hspace{0.5em}} Misdirected correction (disparity widened for already-disadvantaged group)}
\end{table*}
\begin{table*}[ht]
\centering
\caption{Gender fairness metrics across all methods. TPR: True Positive Rate, FPR: False Positive Rate, Pos.\ Rate: Positive classification rate. ``Unknown Gender'' includes students with non-binary identities, unreported gender, and missing data. GetFair reports $\lambda = 0.3$.}
\label{tab:gender-all-methods}
\begin{tabular}{@{}lrrr|rrr|rrr@{}}
\toprule
& \multicolumn{3}{c}{\textbf{Female}} 
& \multicolumn{3}{c}{\textbf{Male}} 
& \multicolumn{3}{c}{\textbf{Unknown Gender}} \\
\cmidrule(lr){2-4} \cmidrule(lr){5-7} \cmidrule(lr){8-10}
\textbf{Method} & TPR & FPR & Pos.R 
    & TPR & FPR & Pos.R 
    & TPR & FPR & Pos.R \\
\midrule
Base        
    & 0.931 & 0.476 & 0.832 
    & 0.876 & 0.321 & 0.706 
    & 0.667 & 0.125 & 0.357 \\
Calibrated  
    & 0.966 & 0.561 & 0.878 
    & 0.931 & 0.406 & 0.770 
    & 0.667 & 0.292 & 0.452 \\
\midrule
GetFair     
    & 0.977 & 0.609 & 0.897 
    & 0.949 & 0.450 & 0.796 
    & \imp{0.722} & \dgr{0.417} & 0.548 \\
Decoupled   
    & 0.966 & 0.561 & 0.878 
    & 0.931 & 0.406 & 0.770 
    & 0.667 & 0.292 & 0.452 \\
Reject Option 
    & 0.941 & 0.496 & 0.845 
    & 0.894 & 0.341 & 0.724 
    & \imp{0.722} & \dgr{0.417} & 0.548 \\
MBS         
    & \imp{0.998} & \dgr{0.837} & 0.963 
    & \imp{0.995} & \dgr{0.734} & 0.914 
    & \imp{1.000} & \dgr{0.667} & 0.810 \\
Bias Mitigation 
    & 0.966 & 0.561 & 0.878 
    & 0.931 & 0.406 & 0.770 
    & 0.667 & 0.292 & 0.452 \\
GBCP        
    & \dgr{0.593} & 0.121 & 0.491 
    & \dgr{0.635} & 0.118 & 0.476 
    & \imp{0.722} & 0.333 & 0.500 \\
\bottomrule
\end{tabular}

\smallskip
\noindent{\footnotesize
\colorbox{green!15}{\strut\hspace{0.5em}} Improvement from calibrated baseline ($\Delta$TPR${>}$0.03 or $\Delta$FPR${>}$0.05) \quad 
\colorbox{red!15}{\strut\hspace{0.5em}} Degradation from calibrated baseline \quad 
\colorbox{orange!20}{\strut\hspace{0.5em}} Misdirected correction (disparity widened for already-disadvantaged group)}
\end{table*}

\begin{table*}[ht]
\centering
\caption{Residency status fairness metrics across all methods. TPR: True Positive Rate, FPR: False Positive Rate, Pos.\ Rate: Positive classification rate. GetFair reports $\lambda = 0.3$.}
\label{tab:residency-all-methods}
\begin{tabular}{@{}lrrr|rrr@{}}
\toprule
& \multicolumn{3}{c}{\textbf{Domestic}} 
& \multicolumn{3}{c}{\textbf{International}} \\
\cmidrule(lr){2-4} \cmidrule(lr){5-7}
\textbf{Method} & TPR & FPR & Pos.\ Rate 
    & TPR & FPR & Pos.\ Rate \\
\midrule
Base        
    & 0.875 & 0.374 & 0.704 
    & 0.937 & 0.413 & 0.856 \\
Calibrated  
    & 0.921 & 0.438 & 0.756 
    & 0.981 & 0.563 & 0.916 \\
\midrule
GetFair     
    & 0.948 & 0.499 & 0.794 
    & 0.981 & 0.564 & 0.916 \\
Decoupled   
    & 0.921 & 0.438 & 0.756 
    & 0.981 & 0.563 & 0.916 \\
Reject Option 
    & 0.883 & \imp{0.384} & 0.713 
    & 0.983 & \mis{0.582} & \mis{0.921} \\
MBS         
    & 0.921 & 0.438 & 0.756 
    & 0.978 & 0.551 & 0.912 \\
Bias Mitigation 
    & 0.921 & 0.438 & 0.756 
    & \mis{0.990} & \mis{0.773} & \mis{0.956} \\
GBCP        
    & \dgr{0.629} & 0.164 & 0.470 
    & \dgr{0.500} & 0.087 & 0.436 \\
\bottomrule
\end{tabular}

\smallskip
\noindent{\footnotesize 
\colorbox{green!15}{\strut\hspace{0.5em}} Improvement from calibrated baseline ($\Delta$TPR${>}$0.03 or $\Delta$FPR${>}$0.05) \quad \colorbox{red!15}{\strut\hspace{0.5em}} Degradation from calibrated baseline \quad \colorbox{orange!20}{\strut\hspace{0.5em}} Misdirected correction (disparity widened for already-disadvantaged group)}
\end{table*}

\begin{table*}[!t]
\centering
\footnotesize
\setlength{\tabcolsep}{3pt}
\caption{Summary of fairness--accuracy trade-offs across evaluated methods.}
\label{tab:method_summary}
\begin{tabular}{p{1.6cm}p{3.5cm}p{5cm}}
\toprule
\textbf{Method} & \textbf{Fairness Gains} & \textbf{Trade-offs / Limitations} \\
\midrule
Calibration & 
Domestic TPR +0.046; gender TPR gap narrowed (0.055$\to$0.035); pos.\ rate convergence improved. & 
Residual FPR gaps (gender: 0.406 vs.\ 0.561); modest small-group improvements. \\
\addlinespace[2pt]
GetFair ($\lambda{=}0.3$) & 
Domestic TPR 0.921$\to$0.948; under-25 TPR 0.938$\to$0.958; age pos.\ rate range narrowed by 0.02. & 
Single threshold cannot equalize heterogeneous base rates; gender DP immovable (0.628--0.634); $\lambda \geq 0.7$ converged to baseline. \\
\addlinespace[2pt]
Decoupled Classifiers & 
High TPR for large groups (age 25--35: 0.979; female: 0.966). & 
Unknown-gender TPR=0.667; defaults to negative class for small groups; reproduced calibrated baseline. \\
\addlinespace[2pt]
Reject Option &
Domestic FPR reduced (0.438$\to$0.384); modest age improvements. &
Misdirected residency correction: intl.\ pos.\ rate 0.916$\to$0.921, domestic 0.756$\to$0.713 (gap widened); group size unreliable proxy for disadvantage. \\
\addlinespace[2pt]
MBS & 
Gender EOp $\leq$0.05 met (F: 0.998, M: 0.995, N: 1.000); residency $\leq$0.10 met. & 
Gender FPRs surged (F: 0.561$\to$0.837, M: 0.406$\to$0.734); $\leq$0.05 infeasible for residency; strict enforcement produced all-zero predictions. \\
\addlinespace[2pt]
Bias Mitigation & 
Unknown age TPR 0.937$\to$0.975; modest smallest-group corrections. & 
Misdirected residency correction: intl.\ FPR 0.563$\to$0.773, pos.\ rate 0.916$\to$0.956 (gap widened). Gender corrections near baseline. \\
\addlinespace[2pt]
GBCP & 
Tightest statistical parity (pos.\ rates 0.47--0.50); FPR range 0.087--0.333. & 
Severe TPR drops (0.30--0.37 from baseline); unknown-gender TPR rose modestly (0.667$\to$0.722); ${\sim}$50\% flag rate exceeds advising capacity. \\
\bottomrule
\end{tabular}
\end{table*}

\subsection{Base Model}
We report discriminative ability with respect to the at-risk class, aligning model evaluation with the intervention goals of EWS. The intake-only base model achieved an AUC of 0.860 (95\% CI: 0.856–0.864), and the calibrated model achieved 0.863 (95\% CI: 0.859–0.867, confidence intervals were computed using 5,000 bootstrap resamples of the test set). These values indicate moderate discriminatory capacity given the absence of in-term academic performance signals, consistent with prior work in early risk stratification. Our analysis therefore focuses on how post-hoc fairness interventions redistribute errors and support opportunities across student groups, rather than attempting to further optimize predictive performance in this constrained setting.
In the base model, we observed substantial disparities across groups. For example, 85.6\% of international students were predicted to succeed compared to only 70.4\% of domestic students—a gap of 15.2 percentage points. Similarly, among students who actually succeeded, the model correctly identified 95.3\% of students aged 25--35 but only 88.3\% of students under 25. These gaps indicate that the model treats different demographic groups quite differently.
Tables \ref{tab:age-all-methods}, \ref{tab:gender-all-methods}, and \ref{tab:residency-all-methods} show fairness metrics for the base model, calibrated model, and all post-hoc interventions across age, gender, and residency status.
Post-hoc calibration narrowed all three fairness metrics without eliminating any of them (Tables~\ref{tab:age-all-methods}--\ref{tab:residency-all-methods}): the gender TPR gap fell from 0.055 to 0.035, and positive rates moved toward parity for domestic and under-25 students. But the gaps did not close. The FPR gap between domestic and international students even widened slightly. Calibration reduces group-level disparities; it does not reach the data imbalances or the structural sources of bias built into institutional practice.

\subsection{Post-Hoc Interventions}
\subsubsection{GetFair: Generalized Fairness Tuning}\label{sec:get-fair}

GetFair \cite{sikdar_getfair_2022} tunes a single global decision threshold by optimizing a reward function that weights fairness against accuracy: $\text{Reward}=\lambda\cdot\text{Fairness}+(1-\lambda)\cdot\text{Accuracy}$, where higher $\lambda$ prioritizes fairness. We evaluated the method across $\lambda$={0.3,0.5,0.7,1.0}, optimizing for demographic parity separately for each sensitive attribute. Table~\ref{tab:age-all-methods} reports results at $\lambda$=0.3 (the most fairness-aggressive configuration that produced operationally distinct results); higher $\lambda$ values converged to near-baseline thresholds, as discussed below.

The central finding is that single-threshold tuning is a weak fairness lever. Even at $\lambda$=0.3, the optimized thresholds (0.39--0.46 depending on attribute) produced only modest departures from the calibrated baseline (threshold $\approx$ 0.50). For age, the under-25 TPR increased from 0.938 to 0.958 and positive rate from 0.787 to 0.819, while the 25--35 group saw TPR rise from 0.979 to 0.988 and positive rate from 0.915 to 0.932. For residency, domestic TPR increased from 0.921 to 0.948 and positive rate from 0.756 to 0.794, while international metrics remained essentially unchanged (TPR: 0.981, positive rate: 0.916). Accuracy remained within 0.3 percentage points of baseline across all configurations (0.837--0.840).

The method's structural limitation is that a single global threshold cannot equalize positive rates across groups with heterogeneous base rates. For gender, the demographic parity difference remained at 0.628--0.634 regardless of $\lambda$ value. At $\lambda \geq 0.7$,
the optimizer converged to thresholds near 0.50, effectively reproducing the calibrated baseline, indicating the method had exhausted its capacity for improvement well before fairness was fully prioritized.

\subsubsection{Decoupled Classifiers}\label{sec:decoupled}
Decoupled classifiers \cite{dwork_decoupled_2018} train group-specific logistic regression models using the base model's probability scores as input. The method essentially reproduced the calibrated baseline for all well-represented groups, with group-specific logistic regressions learning approximately the same decision boundary as global calibration. For unknown gender students ($n=42$), the method showed no improvement (TPR: 0.667, FPR: 0.292, positive rate: 0.452)---the group was too small for the method to learn a calibration function distinct from the pooled baseline. When a group has too few samples (fewer than 20 in our implementation), the method defaults to predicting the negative class, systematically disadvantaging small groups. See Section~\ref{post-hoc descriptions} for full implementation details.

\subsubsection{Reject Option Fairness}\label{sec:reject-option}
The reject option method \cite{kamiran_exploiting_2018} identifies predictions in a region of uncertainty near the decision boundary (controlled by parameter $\theta = 0.6$, corresponding to probabilities between 0.4 and 0.6) and relabels them: students from the ``deprived'' group receive favorable predictions (1), while those from ``favored'' groups receive unfavorable predictions (0). Predictions outside the uncertainty region are unchanged.
 
A key implementation detail shapes these results: the method assigns group privilege based solely on group size, labeling only the single smallest group within each attribute as ``deprived.'' For gender, unknown gender students, the smallest group, received favorable relabeling (positive rate: 0.452 $\to$ 0.548, TPR: 0.667 $\to$ 0.722), though the FPR increase (0.292 $\to$ 0.417) meant more struggling students in this group were missed. The method's single-group ``deprived'' assignment correctly identified the most disadvantaged gender category in this case, but its effectiveness for residency, where the smallest group was already advantaged, demonstrates that group size is an unreliable proxy for disadvantage. As we show in Section~\ref{sec:bias-mitigation}, assigning disadvantage from observed disparity rather than group size redirects this method's correction to domestic students and narrows the gap below its baseline level, confirming that the failure lies in the assignment heuristic rather than the relabeling machinery---though the redirected correction reallocates rather than eliminates error burdens.
 
For age, the method produced modest improvements: under-25 TPR decreased slightly from 0.938 to 0.899 while FPR decreased from 0.430 to 0.360, and domestic FPR decreased from 0.438 to 0.384. For residency, however, the method exhibited the same misdirected correction pattern observed with Bias Mitigation: international students, the smaller group, received ``deprived'' status and favorable relabeling despite already having higher positive rates than domestic students. This widened the residency gap: international positive rate increased from 0.916 to 0.921 and FPR from 0.563 to 0.582, while domestic positive rate decreased from 0.756 to 0.713. Two of our six implementations thus directed corrections at the already-advantaged group for this attribute because both operationalized disadvantage using group size.

\subsubsection{Modification with Bias Scores (MBS)}\label{sec:mbs}
MBS \cite{chen_post-hoc_2024} computes a bias score for each prediction by combining a group membership classifier with the model's confidence, then selectively flips predictions exceeding a learned threshold, subject to a fairness constraint (disparity $\leq$ target) and accuracy maximization. The method optimizes for equalized odds by default, selecting the configuration that achieves the best accuracy while meeting the fairness constraint.

The method's effectiveness is tightly coupled to the feasibility of its fairness constraint given base rate differences. For gender, MBS met the $\leq$ 0.05 constraint on equal opportunity (accuracy: 0.793), producing substantial TPR gains (F: 0.998, M: 0.995, N: 1.000) at the cost of elevated FPRs (F: 0.561 $\rightarrow$ 0.837; M: 0.406 $\rightarrow$ 0.734). For advisors, this trade-off is concrete: meeting the gender equal opportunity constraint means that a substantially larger share of struggling students are classified as ``on track'' and never appear on advising dashboards---the system achieves parity by making failure less visible rather than support more equitable. For age, the selected threshold was so high that metrics were effectively unchanged. For residency, the $\leq$ 0.05 constraint was infeasible---strict enforcement produced all-zero predictions---and cascaded to $\leq$ 0.10 with minimal changes. This sensitivity to constraint specification, where a single parameter choice produces either reasonable corrections or catastrophic failure depending on the attribute's base rate structure, represents a practical barrier for institutional users.

\subsubsection{Bias Mitigation Post-processing}\label{sec:bias-mitigation}
Bias mitigation post-processing \cite{lohia_bias_2018} adapts the Individual and Group Debiasing (IGD) framework for procurement-constrained settings. Since the original method requires counterfactual model queries, which are unavailable under procurement constraints, our adaptation trains a surrogate logistic regression model and uses it to compute individual bias scores: for each student, the method estimates how the prediction would change if the student's demographic attribute were different. Students with high bias scores in the ``unprivileged'' group, identified as the numerically smallest group, receive corrected predictions from the surrogate model with flipped attributes. We ran the full debiasing pipeline separately for each sensitive attribute.

The method's reach was narrow. Of 42,138 held-out test records, it altered 86 predictions for age, 706 for residency, and none at all for gender. For age, those 86 changes were concentrated in the Unknown age group (the smallest), raising its TPR from 0.937 to 0.975 at the cost of a substantially increased FPR (0.371 $\rightarrow$ 0.665): more students in this group are now classified as on track, and correspondingly fewer appear on advising caseloads. Well-represented age groups were untouched. For gender, the method changed no predictions whatsoever; the surrogate found no students whose bias scores and group membership met its correction criteria, and F, M, and unknown-gender metrics all remained at their calibrated values.

The most revealing result occurred for residency status. Our adaptation designates the numerically smaller group as unprivileged and directs favorable corrections toward it. For student type, however, international students, the smaller group, already receive \textit{higher} positive rates than domestic students (0.916 vs.\ 0.756 at baseline). All 706 corrected predictions therefore fell in the international group, and the correction widened rather than narrowed the existing disparity: international FPR increased from 0.563 to 0.773 and positive rate from 0.916 to 0.956, while domestic metrics remained exactly unchanged (TPR: 0.921, FPR: 0.438, positive rate: 0.756). The method ``helped'' the group that was already advantaged on this metric, amplifying the gap it was designed to close. Operationally, the correction made struggling international students \textit{less} visible to advisors: an additional 4\% of international students moved off the flagged list, none of them because their predicted risk had been reassessed on its merits.

\paragraph{The misdirection is structural, not stochastic.}
This result does not depend on how the surrogate model happened to initialize. The method identifies the ``unprivileged'' group by taking the minimum of the group counts, a deterministic function of the dataset rather than of the model. Across 20 random seeds, all 706 flipped predictions fell in the international group in every run, international FPR settled at 0.773 and positive rate at 0.956 in every run, and domestic metrics were untouched in every run (Appendix~\ref{app:seed_variation}). The method misdirects its correction because of what it assumes, not because of how it was initialized.

\paragraph{Misdirected corrections as a structural pattern.}
This is the second of six methods to direct favorable corrections at the already-advantaged group for residency status, reproducing through a different mechanism the pattern produced by Reject Option (Section~\ref{sec:reject-option}). Reject Option assigned ``deprived'' status based on group size; Bias Mitigation identified the ``unprivileged'' group as the numerically smallest. 
In our implementations, both methods used group size to designate disadvantage, without checking whether that designation matched observed disparities. Their shared result illustrates a broader failure mode of this operationalization: when the smaller group has a higher positive rate, directing favourable corrections toward it can widen the existing gap. The sensitivity analysis below tests how changing that designation affects each implementation. Neither group size nor observed positive rate, however, establishes which group should be considered disadvantaged in the institutional context.

\paragraph{Isolating the mechanism.}
To confirm that the misdirection stems from the disadvantage-assignment heuristic rather than the correction machinery, we re-ran both methods with ``deprived'' status assigned by observed disparity---the group with the lowest calibrated positive rate---instead of by group size (Appendix~\ref{app:disadvantage_sensitivity}). The two methods responded differently, and the difference is instructive. Reject Option's correction redirected as intended: relabeling now favored domestic students, narrowing the domestic--international positive-rate gap from 16.1 points at baseline to 8.6, though at the cost of raising domestic FPR from 0.438 to 0.499, rendering more struggling domestic students invisible to advisors. Bias Mitigation, by contrast, went inert: with the majority group designated as deprived, the pipeline flipped zero predictions, returning output bit-identical to the calibrated baseline. Its bias-scoring thresholds and viability checks were implicitly built around correcting a numeric minority, and no relabeling of groups makes it act on a majority. The heuristic, in other words, is not merely a default that institutions could reconfigure with better knowledge; for this method, the equation of disadvantage with numerical minority is embedded in the correction machinery itself. For age and gender, the two assignment rules coincide---the smallest group is also the group with the lowest positive rate---which is precisely why the failure surfaces only for residency: the heuristic misfires exactly when group size and disadvantage diverge, the condition a method without institutional context cannot detect. The alternative assignment rule is itself a proxy, not a determination of which students are disadvantaged. A lower calibrated positive rate does not establish the source of a disparity or identify the correction students need. An institution-informed definition, developed with students and staff and attentive to the conditions shaping academic outcomes, could designate groups differently from either group size or observed positive rates. Our sensitivity analysis therefore shows that the results depend on how disadvantage is operationalized; it does not establish which assignment rule is fair.

\subsubsection{Group-Balanced Conformal Prediction (GBCP)}\label{sec:conformal}
Group-balanced conformal prediction \cite{angelopoulos_gentle_2022} sets group-specific probability thresholds such that positive prediction rates converge to a shared target (0.50). The method achieved the tightest statistical parity: positive rates converged to 0.47--0.50 across attributes. However, TPR reductions of 0.30--0.37 occurred for well-represented groups, with the largest drops for groups whose baselines diverged most from 0.50 (international: 0.981 to 0.500; female: 0.966 to 0.593). Unknown gender students, already near the 0.50 target, saw a modest TPR increase (0.667 to 0.722). This would flag roughly half of all students for support. At a mid-sized college with, e.g., 15 advisors and 3,000 incoming students, that is about 100 students per advisor for proactive outreach in the first weeks of term, which erases any distinction between flagged and unflagged students and makes the system useless as a triage tool. The method achieves parity by equalizing harm rather than benefit.

\subsection{Comparative Effectiveness}\label{sec:comparison}

\begin{table}[t]
\centering
\caption{Number of predictions changed from the calibrated baseline by each method, per sensitive attribute, out of 42,138 held-out test records. Three methods changed no predictions at all for at least one attribute; GBCP changed roughly a third of all predictions; GetFair at high fairness weights ($\lambda \geq 0.7$) changed none for age and residency. Metrics can converge on dashboards while the underlying classifications---and the students advisors see---remain untouched or shift wholesale.}
\label{tab:rows-changed}
\begin{tabular}{@{}lrrr@{}}
\toprule
\textbf{Method} & \textbf{Age} & \textbf{Gender} & \textbf{Residency} \\
\midrule
GetFair ($\lambda=0.3$) & 1{,}189 & 942 & 942 \\
Decoupled Classifiers & 23 & 0 & 0 \\
Reject Option & 1{,}693 & 1{,}688 & 1{,}141 \\
MBS & 0 & 4{,}842 & 76 \\
Bias Mitigation & 86 & 0 & 706 \\
GBCP & 14{,}326 & 14{,}329 & 15{,}485 \\
\bottomrule
\end{tabular}
\end{table}

Table~\ref{tab:method_summary} summarizes key metrics across all six methods, and Table~\ref{tab:rows-changed} reports how many of the 42,138 test-set predictions each method actually changed. The counts alone tell much of the story: three methods left at least one attribute untouched, while GBCP rewrote roughly a third of all classifications. These counts point to a distinction we develop in Section~\ref{fairness_theatre}. A method can fail in two ways. It can change almost nothing: predictions stay in place, the dashboard holds steady, and existing inequities persist. Or it can force the metrics to converge by reclassifying students wholesale, shifting error burdens from one group to another. Only the second is fairness theatre. We reserve the term for methods that improve the dashboard without improving the burdens students carry.

Across the fairness criteria, the trade-offs are consistent. Only GBCP reached tight statistical parity (positive rates 0.47--0.50), and it did so by cutting TPR sharply for every group whose base rate sat far from the target. MBS closed the gender TPR gap only by driving false positive rates up (female FPR 0.561 $\rightarrow$ 0.837)---the mechanical consequence of flipping unsuccessful students' predictions rather than improving the underlying classification. GetFair, Decoupled, Reject Option, and Bias Mitigation all held accuracy within half a point of baseline, and their fairness changes were correspondingly small. GetFair is the telling case: it preserved accuracy precisely because it could not find thresholds that departed from the baseline, so what looks like a clean trade-off is really the absence of one.

\paragraph{Group-Specific Patterns.} International students received higher positive rates than domestic students across all six methods (a gap ranging from 0.04 in MBS to 0.20 in Bias Mitigation, where the misdirected correction widened it). Female students showed consistently higher FPRs than male students. And unknown-gender students received the lowest TPRs across every method that did not collapse into near-uniform classification. The bootstrap intervals (Appendix~\ref{app:bootstrap_ci}) sharpen this last point: unknown-gender TPR falls below both binary-gender estimates in the four methods with a normal error profile, and does not overlap them for three. MBS and GBCP are only apparent exceptions---MBS reaches a higher unknown-gender TPR by flipping the group to an all-positive state, GBCP by dragging female and male TPR down to 0.59--0.63---so the higher number signals the method collapsing, not correcting.

\paragraph{Error Type Profiles.} The six methods sort into three profiles, each reshaping advising work differently. \textit{False-positive-generating methods} (MBS, Bias Mitigation) close TPR gaps by selectively flipping predictions, fixing some false negatives while creating new false positives; caseloads shrink, and struggling students who would once have been flagged now read as ``on track'' and get no outreach. \textit{Sensitivity-reducing methods} (GBCP) cut false positives but generate false negatives in bulk, producing caseloads so large that advisors cannot triage within them. \textit{Baseline-preserving methods} (GetFair, Decoupled) hold the calibrated error profile and redistribute almost nothing---the workflow is untouched, and so are its existing inequities. Reject Option falls between profiles, its error pattern depending on which group its logic marks as disadvantaged.

\section{Discussion} \label{discussion}
The value of our empirical work lies not in discovering that fairness trade-offs exist (impossibility theorems guarantee they must \cite{kleinberg_inherent_2016, chouldechova_fair_2017}) but in demonstrating \textit{which students pay the price} for different fairness choices, \textit{how large} these costs are in practice, and \textit{how implementation decisions can silently determine outcomes}. This section synthesizes four critical themes: the performance of fairness under procurement constraints; the differential impact across student populations; the multi-dimensional nature of fairness-accuracy-interpretability trade-offs; and the limits of our own approach.

\subsection{Procurement and the Performance of Fairness in Educational AI}
\label{sec:accountability_gap}
Prior ethnographic work theorized this dynamic as the ASP-HEI Cycle (Section~\ref{sec:background}), in which financial pressures drive institutions into vendor dependence that in turn exacerbates existing inequities \cite{mcconvey_this_2024}. Our findings show what happens at one node of that cycle: an institution, already locked into a vendor's system, tries to govern the fairness of outputs it can see but cannot interrogate. 
Institutions reliant on vendor-controlled models cannot query, retrain, or meaningfully interrogate system design, leaving surface-level post-hoc adjustments as one of the only available levers \cite{veale_fairness_2018}. Fairness here is not a property of the algorithm. It is something vendors, institutional staff, and students have to accomplish together---and procurement contracts make that
joint work nearly impossible by cutting off the information and the levers it requires. Three things box in what the institution can do. It can only act after the model has already encoded discriminatory patterns, not before. The vendor knows the model and its training data; the institution does not. And the procurement rules it operates under were written for buying goods and services, not for governing algorithms that keep changing \cite{mulligan_procurement_2019}.

\label{sec:solutionism_trap}
Within these constraints, post-hoc methods risk what \citet{selbst_fairness_2019} call the ``solutionism trap'', treating structural inequities as technical imbalances to optimize. We see three failure modes. First, methods that aggressively pursue parity can produce operationally untenable results: GBCP would flag approximately half of all students for support, overwhelming advising capacity (Section~\ref{sec:conformal}). Second, methods that preserve accuracy leave disparities largely intact: GetFair and Decoupled Classifiers both produced near-baseline metrics for well-represented groups while offering no meaningful correction for small categories. Third, methods can actively worsen the disparities they are designed to address: both Reject Option and Bias Mitigation directed favorable corrections toward international students, who already had higher positive rates, widening the residency gap rather than closing it (Sections~\ref{sec:reject-option},~\ref{sec:bias-mitigation}). Two of our six implementations, built on different techniques, produced this pattern because both designated disadvantage by group size. Any method that equates group size with disadvantage will point its correction the wrong way whenever the smaller group is not the worse-off one. Sensitivity analysis
(Section~\ref{sec:bias-mitigation}) confirms it: reassigning disadvantage by observed disparity redirects Reject Option's correction but leaves Bias Mitigation unchanged. For our Bias Mitigation adaptation, ``smaller group means disadvantaged'' is not merely a configurable label but an assumption built into the adapted correction pipeline.

We interpret these failure modes through the ASP-HEI cycle \cite{mcconvey_this_2024}, while acknowledging that each also has a plain technical explanation that needs no structural account: impossibility constraints, methods with little room to adjust, crude group-size heuristics. What the structural reading adds is the connection between them. The parity these methods reach only by flagging half the student body reflects the resource scarcity that drives adoption in the first place; accuracy-preserving inertia reflects the vendor dependency that forecloses any real intervention; and misdirected corrections reflect the cycle's core prediction, that adopting these systems reproduces the inequities already in place. The two kinds of explanation are not rivals. The technical one says what each method did; the structural one says why institutions were using such methods at all.

EWS are built to predict, but institutions adopt them to intervene, and the two are not the same thing. \citet{liu_bridging_2025} argue that a prediction system in a social setting is really a policy intervention: its value depends on the institution's context, the actions available, and the people involved, not on the score itself. Post-hoc fairness methods make this worse. They fine-tune who receives a prediction, while whether that prediction leads to any help depends on advising capacity, the kinds of support on offer, and infrastructure, none of which the method touches.

\label{fairness_theatre}
This is what we call \emph{fairness theatre}: a method's outputs look parity-seeking on the dashboard while the underlying inequities stay put or grow. MBS is the clearest case. To close the gender gap in equal opportunity, it flipped predictions until the TPR gaps nearly vanished but it drove false positive rates up sharply to do it (Section~\ref{sec:mbs}). On the dashboard, that reads as improved equal opportunity. On the ground, it means more struggling students get labeled ``on track'', more students in the position of the international student from Section~\ref{sec:background}, who never appears on any list. The metric improves because of his absence, not despite it. When we applied the same strict constraint to residency, MBS returned all-zero predictions until we relaxed it: one parameter choice that produces sensible corrections for one attribute and catastrophic failure for another, depending entirely on the group base rates. A method that behaves this way needs exactly
the contextual judgment that procurement takes away. Fairness theatre also serves a purpose within the ASP-HEI cycle. It lets an institution show accountability to outside stakeholders and hit the performance metrics that
financial-sustainability pressures demand \cite{mcconvey_this_2024}, all while the conditions actually producing the inequity stay untouched.

These findings put empirical weight behind the theoretical limits we laid out in Section~\ref{sec:limitations}. Post-hoc interventions are a case of what \citet{green_algorithmic_2020} calls algorithmic formalism's ``narrowing of vision'', and they bear out \citet{kasirzadeh_algorithmic_2022}'s point that locally redistributive adjustments can carry ``negligible social significance'' when the structural injustice goes unaddressed. These methods can only ever manage symptoms, because procurement and limited institutional capacity block the institution from touching the features, training data, and outcome definitions where the real inequities are built in.

Interpretability pays a price too. When MBS flips predictions to satisfy equal opportunity, two students with identical risk scores can end up with different support, decided only by which group's threshold applied---an opaque correction stacked on an already-opaque vendor output. This deepens what prior work describes as the automation of the faculty--student relationship \cite{mcconvey_this_2024}: an advisor acting on an adjusted prediction is one more step removed from the judgment that discretion used to allow, and the normative choices baked into a method's defaults, Reject Option treating group size as disadvantage for instance, stay invisible to anyone who sees only outputs.

\subsection{Intersectional Erasure in Vendor-Constrained Fairness}\label{sec:representation_paradox}
Single-axis fairness methods, those that adjust for one attribute at a time, erase students at the margins. In the methods that behaved normally, unknown-gender students had the lowest TPRs, while the binary gender categories exceeded 0.93 in most methods. The two apparent exceptions do not count: MBS pushed unknown-gender TPR to a perfect 1.000 by classifying nearly everyone as successful, and GBCP's small gain (0.667 to 0.722) came as a side effect of reclassifying a third of all students (Sections~\ref{sec:mbs},~\ref{sec:conformal}). No method actually served this group better. And because the gap survived six very different optimization strategies, the failure is not a quirk of any one method. Adjusting one attribute at a time cannot reach students who are marginalized along several at once \cite{crenshaw_demarginalizing_1989, hanna_towards_2020}.

Part of the problem is what these methods take a demographic category to be. They treat it as a fixed attribute to balance. But the disparities in our data come from institutional processes: admissions policies that favour
domestic students, advising built around traditional-age students, data collection that assumes gender is binary \cite{hanna_towards_2020}. Bias Mitigation shows the cost of ignoring this (Section~\ref{sec:bias-mitigation}). It took group size as a proxy for disadvantage and balanced the category without ever asking what produced the disparity. Small groups make this worse: methods that need large samples fail first for the students most exposed to discrimination.

Prior work holds that adopting these systems drives more data collection and surveillance \cite{mcconvey_this_2024}. Our findings show who that data actually serves. Students in large, well-represented categories, the ones the data sees most clearly, are the ones fairness methods help most. The students most at risk of algorithmic harm are the ones the data barely records: gender-diverse students collapsed into ``Unknown,'' students whose disadvantages compound across categories no single-axis method can see. More
surveillance does not buy more fairness. It buys more precision for the already-visible.

\subsection{Implications for Institutional Practice}
\label{implications and limitations}
Selecting among post-hoc methods is not merely technical: it encodes normative choices about which errors are acceptable and for whom \cite{jarrahi_principles_2023}. More fundamentally, institutions choosing among these methods sit at the far downstream end of the cycle. The upstream conditions---vendor dependence, scarce resources, opaque contracts---have already decided which options are on the table and which costs the institution can absorb \cite{mcconvey_this_2024}. Meaningful intervention therefore
requires acting at several points at once.

\subsubsection{From Post-Hoc Diagnosis to Institutional Action}
Meaningful fairness in educational AI requires coordinated interventions. Each of the following addresses a different node of the cycle theorized in \cite{mcconvey_this_2024}:

\textbf{Procurement reform with accountability mechanisms:} Contracts should specify not only fairness thresholds but transparency in model development (training-data demographics, feature decisions) and ongoing accountability (regular audits shared with student representatives, contractual remedies when disparities exceed agreed thresholds, and rights to retrain when institutional contexts change).

\textbf{Participatory governance structures:} Drawing on participatory AI in public-sector contexts \cite{hanna_towards_2020, chui_towards_2025}, institutions could establish student advisory boards with authority over fairness criteria. Our findings make concrete the trade-offs such boards would weigh---FPR costs for equal opportunity, feasible flagging rates, whether misdirecting methods should be deployed at all---rendering visible choices currently hidden from institutional users.

\textbf{Data governance reform:} Participatory data practices that let students self-define identity categories, set consent preferences, and contest classifications would address the intersectional invisibility we documented, responding to critiques that fairness methods treat demographic attributes as static rather than relational \cite{hanna_towards_2020}.

\textbf{Capacity-aligned fairness planning:} Because some fairness-optimized configurations exceed advising capacity or hollow out prediction meaningfulness, fairness interventions must be coupled with resource decisions. Liu et al.\ \cite{liu_bridging_2025} show that in resource-constrained settings, expanding intervention capacity can yield greater welfare gains than optimizing prediction targeting \cite{liu_actionability_2024}.

These interventions require resources and sustained institutional commitment beyond what post-hoc methods can provide, and the constraints shaping them (procurement restrictions, capacity limits, intersectional invisibility) are structural features of public-sector AI procurement rather than peculiarities of our site.

\subsection{Limitations, Reflexivity, and Future Directions}\label{sec:limitations-directions}
Our analysis embodies a tension identified throughout the critical fairness literature: we use fairness metrics to evaluate interventions in a domain where those metrics cannot capture justice \cite{jacobs_measurement_2021, vethman_fairness_2025}. We justify this pragmatically—institutions are deploying these methods and need evidence about what they can and cannot achieve under realistic constraints—but the contradiction should be named. Our contribution is showing that even when post-hoc methods optimize multiple fairness definitions simultaneously, they cannot substitute for structural reform \cite{selbst_fairness_2019}. Our findings should therefore be understood as \textit{motivating} rather than \textit{completing} fairness work.

Several specific limitations shape our conclusions' scope. As we note in Section~\ref{Data and Modeling}, our binary outcome conflates financial withdrawal, program mismatch, caregiving, and academic failure into a single ``unsuccessful'' label \cite{baker_algorithmic_2022}. While we evaluate fairness under realistic institutional constraints, we cannot determine whether international students' higher positive rates (0.79--0.96) reflect algorithmic bias, admissions selection effects, or differential support structures \cite{kasirzadeh_algorithmic_2022}. Our analysis draws on one Ontario college, limiting generalizability—particularly post-COVID and post-2024 visa caps \cite{crawley_ontarios_2023}. Our demographic categories (age, gender, residency) lack race, ethnicity, disability, and socioeconomic data, and our single-axis framework could not assess intersectional subgroups \cite{crenshaw_demarginalizing_1989,hanna_towards_2020}.

More fundamentally, our metrics cannot show how students experience being flagged as ``at-risk''; which institutional decisions produce differential success rates; whether statistical convergence constitutes fairness in context; or how intersectional identities compound marginalization beyond single-axis analysis. Our classification framework also ignores how ``at-risk'' labeling affects students' academic identity and belonging \cite{andalibi_conceptualizing_2023}, and we simulate interventions on historical data without observing student and advisor responses, self-fulfilling prophecy effects, or how capacity constraints shape actual support delivery \cite{liu_delayed_2018}.
We do not claim that our exact numerical trade-offs generalize to all educational or public-sector deployments; rather, we expect the patterns we document (e.g., instability for small groups, tension between parity and capacity, misdirected corrections when methods assume disadvantage without checking, sensitivity of fairness constraints to base rate structures) to recur in vendor-controlled, procurement-constrained systems.

\section{Conclusion}
Using a held-out test set of 42,138 student records drawn from a 168,550-record modeling cohort, we stress-tested six post-hoc fairness interventions under the constraints procurement actually imposes. No method delivered consistent equity gains. Improvements for one group came with new gaps for another, well-represented students benefited most, and marginalized ones stayed disadvantaged. Beyond these expected trade-offs, we found three patterns that should trouble anyone deploying these methods: an implementation can operationalize disadvantage in a way that contradicts observed disparities, as two of ours did for residency; a single constraint that corrects reasonably for one attribute can fail catastrophically for another; and an implementation choice invisible to the institution can decide whether a method helps, does nothing, or does harm.
These are not independent obstacles. They are linked stages of the same structural dynamic \cite{mcconvey_this_2024}: financial pressures produce the vendor dependencies that foreclose real intervention, the scarce advising capacity that makes parity-seeking methods unworkable, and the data practices that render students at the margins invisible. A post-hoc method acting at a single downstream point of that cycle cannot reach the conditions that generate the disparities it is asked to correct. Fairness, on this account, is not a one-time optimization but ongoing work of negotiation and judgment, distributed across actors with deeply unequal power to carry it out.
This is where our contribution to CSCW sits. We reframe post-hoc fairness in procured AI as a problem of cooperative work rather than a property of a model; we introduce error-type profiling to translate fairness metrics into the caseloads advisors actually manage; and we document fairness theatre empirically, showing how adjustments can satisfy dashboard metrics while the burdens students carry stay put or grow. Prior work theorized how algorithm adoption consolidates institutional power \cite{mcconvey_this_2024}. Our findings show that fairness interventions, confined to the narrow space vendor control permits, can become instruments of that consolidation rather than checks on it. Closing the gap will take more than better tools. It will take governance that reclaims oversight of algorithmic design, aligns institutional capacity with the support students need, and refuses to accept the appearance of fairness in place of its substance.

\bibliographystyle{ACM-Reference-Format}
\bibliography{references}
\appendix
\section{Appendix} \label{appendix}
\subsection{Seed Variation for Bias Mitigation Post-processing}\label{app:seed_variation}

The Bias Mitigation pipeline (Section~\ref{sec:bias-mitigation}) trains a surrogate logistic regression whose solver (\texttt{saga}) is stochastic. To confirm that our reported results are not artifacts of a particular initialization, we re-ran the full pipeline across 20 random seeds (0--19), recomputing all group-level metrics each time. Table~\ref{tab:seed-variation} reports the results.

Variation is negligible. For gender and residency, every group-level metric is invariant across all 20 seeds to within floating-point precision, and the number of altered predictions is identical in every run (0 for gender, 706 for residency). For age, the only measurable variation occurs in the Unknown category ($n=778$), the smallest group and the one the method treats as unprivileged for that attribute; its FPR varies within a range of 0.005 and its positive rate within 0.002, while its TPR is invariant. The number of altered predictions for age is 86 in 19 of 20 runs and 87 in one.

The residency finding is fully deterministic across seeds. In all 20 runs, every one of the 706 corrected predictions fell in the international group, international FPR settled at 0.773 and positive rate at 0.956, and no domestic prediction was altered. This is expected: the method selects its ``unprivileged'' group by taking the minimum of the group counts, a property of the dataset rather than of the model, so the direction of the correction cannot vary with the seed.

\begin{table*}[t]
\centering
\footnotesize
\caption{Seed variation for Bias Mitigation post-processing across 20 random seeds (0--19). Metrics with a range of zero are invariant to within floating-point precision. The only measurable variation across all three attributes occurs in the Unknown age category, the smallest group in the dataset.}
\label{tab:seed-variation}
\begin{tabular}{@{}llrrrr@{}}
\toprule
\textbf{Attribute} & \textbf{Group} & \textbf{TPR (range)} & \textbf{FPR (range)} & \textbf{Pos.\ Rate (range)} & \textbf{$n$} \\
\midrule
\multirow{4}{*}{Age}
  & Under 25 & 0.938 (0) & 0.429 (0) & 0.786 (0) & 29{,}114 \\
  & 25 to 35 & 0.979 (0) & 0.621 (0) & 0.915 (0) & 8{,}078 \\
  & Over 35  & 0.966 (0) & 0.668 (0) & 0.912 (0) & 4{,}168 \\
  & Unknown  & 0.975 (0) & 0.666 (0.005) & 0.887 (0.002) & 778 \\
\midrule
\multirow{3}{*}{Gender}
  & Female   & 0.966 (0) & 0.561 (0) & 0.878 (0) & 20{,}830 \\
  & Male     & 0.931 (0) & 0.406 (0) & 0.770 (0) & 21{,}266 \\
  & Unknown  & 0.667 (0) & 0.292 (0) & 0.452 (0) & 42 \\
\midrule
\multirow{2}{*}{Residency}
  & Domestic      & 0.921 (0) & 0.438 (0) & 0.756 (0) & 24{,}400 \\
  & International & 0.990 (0) & 0.773 (0) & 0.956 (0) & 17{,}738 \\
\bottomrule
\end{tabular}
\end{table*}

\subsection{Bootstrap Confidence Intervals}\label{app:bootstrap_ci}
We computed 95\% percentile confidence intervals for every group-level metric via stratified bootstrap over the evaluation set (5,000 replicates per method--attribute--group--metric combination, with each method's learned parameters held fixed so that the intervals capture evaluation uncertainty rather than retraining variability). The full set of 216 intervals is available in our replication materials. Table~\ref{tab:unknown-gender-ci} reports intervals for the unknown-gender category, the group whose small size ($n=42$ in the test set) makes uncertainty most consequential for interpretation.

\begin{table}[h]
\centering
\caption{}
\label{tab:unknown-gender-ci}
\begin{tabular}{@{}lrrr@{}}
\toprule
\textbf{Method} & \textbf{TPR} & \textbf{FPR} & \textbf{Pos.\ Rate} \\
\midrule
Base & 0.667 [0.44, 0.88] & 0.125 [0.00, 0.27] & 0.357 [0.21, 0.50] \\
Calibrated & 0.667 [0.43, 0.88] & 0.292 [0.12, 0.48] & 0.452 [0.31, 0.60] \\
\midrule
GetFair & 0.722 [0.50, 0.92] & 0.417 [0.22, 0.62] & 0.548 [0.40, 0.69] \\
Decoupled & 0.667 [0.44, 0.88] & 0.292 [0.12, 0.48] & 0.452 [0.31, 0.60] \\
Reject Option & 0.722 [0.50, 0.93] & 0.417 [0.23, 0.62] & 0.548 [0.40, 0.69] \\
MBS & 1.000 [1.00, 1.00] & 0.667 [0.47, 0.85] & 0.810 [0.69, 0.93] \\
Bias Mitigation & 0.667 [0.44, 0.88] & 0.292 [0.12, 0.48] & 0.452 [0.31, 0.60] \\
GBCP & 0.722 [0.50, 0.93] & 0.333 [0.16, 0.52] & 0.500 [0.36, 0.64] \\
\bottomrule
\end{tabular}
\end{table}

\subsection{Sensitivity of Disadvantage Assignment}\label{app:disadvantage_sensitivity}
Reject Option and Bias Mitigation both identify the group to correct using group size. We re-ran both methods with an alternative rule assigning ``deprived'' status to the group with the lowest calibrated positive rate, holding every other parameter fixed. Under the original rule, the reproduction is exact: row-change counts (Bias Mitigation 86/0/706; Reject Option 1{,}693/1{,}688/1{,}141 across age/gender/residency) and all group-level metrics match the values reported in Tables~\ref{tab:age-all-methods}--\ref{tab:residency-all-methods}. Table~\ref{tab:disadvantage-sensitivity} compares the two rules for residency status, the one attribute where they designate different groups.

\begin{table}[h]
\centering
\footnotesize
\caption{Sensitivity of disadvantage assignment for residency status. Rule A (original): the smallest group is designated ``deprived.'' Rule B: the group with the lowest calibrated positive rate is designated ``deprived.'' All other parameters held fixed. Reject Option's correction redirects under Rule B; Bias Mitigation alters zero predictions.}
\label{tab:disadvantage-sensitivity}
\begin{tabular}{@{}llrrrr|rrrr@{}}
\toprule
 & & \multicolumn{4}{c}{\textbf{A: smallest = deprived}} & \multicolumn{4}{c}{\textbf{B: lowest pos.\ rate = deprived}} \\
\cmidrule(lr){3-6} \cmidrule(lr){7-10}
\textbf{Method} & \textbf{Group} & TPR & FPR & Pos.R & $n$ chg. & TPR & FPR & Pos.R & $n$ chg. \\
\midrule
Reject Option & domestic & 0.883 & 0.384 & 0.712 & \multirow{2}{*}{1{,}141} & 0.948 & 0.499 & 0.794 & \multirow{2}{*}{1{,}579} \\
 & international & 0.983 & 0.582 & 0.922 & & 0.957 & 0.462 & 0.881 & \\
\addlinespace
Bias Mitigation & domestic & 0.921 & 0.438 & 0.756 & \multirow{2}{*}{706} & \multicolumn{4}{c}{\multirow{2}{*}{no change (0 predictions altered)}} \\
 & international & 0.990 & 0.773 & 0.956 & & \multicolumn{4}{c}{} \\
\bottomrule
\end{tabular}
\end{table}

\subsection{Detailed Descriptions of Fairness Metrics}\label{fairness metrics}
\textbf{Demographic Parity} (DP), also called \textit{Statistical Parity} or linked to disparate impact, requires that the probability of a positive prediction (here: predicted “successful”) be equal across protected groups. In our context, this is the \textbf{positive classification rate}—the share of students from each group not flagged for support. Large disparities raise concerns about equal treatment, with certain groups disproportionately subjected to additional monitoring. Approaches to achieving DP include statistical parity optimization \cite{dwork_decoupled_2018} and discrimination minimization \cite{kamiran_exploiting_2018,lohia_bias_2018,sikdar_getfair_2022}.

\textbf{Equal Opportunity} (EOp) focuses on group differences in the \textbf{true positive rate} (TPR): the proportion of students who truly succeed and are correctly predicted to succeed. Lower TPR for a group implies over-intervention—misclassifying successful students as “at risk,” potentially stigmatizing them or undermining trust \cite{chen_post-hoc_2024,sikdar_getfair_2022}.

\textbf{Equalized Odds} (EO) extends EOp by requiring both TPR and the \textbf{false positive rate} (FPR) to be equal across groups. High FPR for a group means struggling students are incorrectly predicted to succeed and therefore miss out on needed support \cite{chen_post-hoc_2024,lohia_bias_2018,sikdar_getfair_2022}.

Finally, \textbf{ranking-based fairness metrics} such as xAUC and Pairwise Ranking Fairness (PRF) \cite{cui_towards_2021} are designed for scenarios where score ordering matters (e.g., admissions ranking). These are less applicable to our binary classification setting, where the decision is framed as “successful” vs.\ “unsuccessful” and used to allocate support resources.
\subsection{Detailed Descriptions of Post-Hoc Techniques} \label{post-hoc descriptions}
\paragraph{GetFair: Generalized Fairness Tuning of Classification Models} \label{getfair}
To assess and mitigate disparities in model predictions, we implemented a post-hoc threshold tuning procedure inspired by the GetFair framework proposed by Sikdar et al.\ \cite{sikdar_getfair_2022}. Rather than retraining the classifier, this method iteratively adjusts the model's decision threshold, optimizing for a weighted combination of fairness and accuracy.
 
We applied threshold tuning separately for each sensitive attribute: age group, gender, and residency status. For each attribute, we converted the calibrated risk scores to probability-of-success values ($P(\text{success}) = 1 - \text{risk score}$) and applied binary classification at the optimized threshold. The tuning process was guided by a reward function:
 
\[
\text{Reward} = \lambda \cdot \text{Fairness} + (1 - \lambda) \cdot \text{Accuracy}
\]
 
where $\lambda$ controls the trade-off between the fairness metric (demographic parity difference, computed using \texttt{fairlearn.metrics}) and predictive accuracy. We evaluated four $\lambda$ values: 0.3, 0.5, 0.7, and 1.0. Each configuration was run five times with different random seeds (episodes = 500), and we used the median threshold for stability.
 
The tuning algorithm performs a random search over the threshold space: at each episode, it randomly perturbs the current threshold, evaluates the resulting reward, and updates via a policy-gradient-inspired step. Unlike the original GetFair paper, which optimizes model hyperparameters using a meta-optimizer network, our implementation tunes only the decision threshold, a deliberate simplification that mirrors the limited intervention space available under procurement constraints, where institutions can adjust classification boundaries but not model parameters.
 
We report results at $\lambda = 0.3$ in the main tables, as this was the most fairness-aggressive configuration that produced operationally distinct results. At $\lambda \geq 0.7$, the optimizer converged to thresholds near the calibrated baseline ($\approx$0.50), as the single-threshold architecture cannot find configurations that improve demographic parity without degrading accuracy for at least one group, a structural limitation rather than an optimization failure.

\paragraph{Decoupled Classifiers for Group-Fair and Efficient Machine Learning} \label{decoupled}
\textit{Note on implementation:} The original decoupled classifiers method \cite{dwork_decoupled_2018} trains separate models for each group from scratch, requiring full retraining access. Our implementation adapts this approach for post-hoc use under procurement constraints: we train group-specific logistic regression models using only the base model's predicted probability scores as input, treating them as a calibration layer rather than replacing the base model.

Next, we applied the decoupled classifiers method introduced by Dwork et al. \cite{dwork_decoupled_2018}. This post-hoc approach trains a separate classifier for each group defined by a single sensitive attribute using the model’s predicted probability score as input. For each attribute, we partitioned the data by group and fit group-specific logistic regression models. These models were calibrated using Platt scaling, a calibration technique that adjusts probability estimates, to ensure reliable probabilistic predictions across groups. We then used the resulting decoupled classifiers to generate post-hoc predictions and computed group-wise fairness metrics, including true positive rate (equal opportunity), false positive rate (equalized odds), and positive prediction rate (statistical parity). We compared these metrics across groups within each attribute to assess residual disparities.

To select the best-performing models, we introduced a joint loss function that balances predictive accuracy with fairness disparities. Specifically, we evaluated combinations of group-specific models using a tunable parameter $\lambda$ to control the tradeoff between classification loss and group-level fairness deviations. This enabled us to identify models that offered improved fairness outcomes without unacceptable losses in predictive performance.

In addition, we incorporated transfer learning to improve performance for underrepresented groups. Each group-specific model was trained using a weighted combination of in-group and out-group examples, modulated by a parameter $\theta$. Smaller values of $\theta$ reduce the influence of out-group data, preserving group specificity, while larger values improve generalization for low-resource groups. We used grid search to jointly optimize $\lambda$ and $\theta$ for each sensitive attribute, selecting configurations that minimized joint loss.

We did not attempt to simultaneously balance fairness across all sensitive features (e.g., across intersections of age, gender, and residency status), as the decoupled classifier method is designed for disjoint group partitions based on a single attribute. Extending it to handle multi-attribute intersectional groups would require exponentially more group-specific models, increasing the risk of overfitting and limiting reliability in underrepresented groups. Instead, we assessed fairness separately across each attribute to identify and interpret disparities in a tractable and interpretable manner.
\paragraph{Exploiting Reject Option in Classification for Social Discrimination Control} \label{exploiting}
The next method we explored was the reject option fairness method proposed by Kamiran et al. \cite{kamiran_exploiting_2018}. For each sensitive attribute, we identified instances whose predicted probabilities fell within a threshold-defined region of uncertainty around the decision boundary. These instances were considered ``rejects'' and were relabeled based on group membership: individuals from disadvantaged groups were assigned favorable outcomes, while those from advantaged groups received unfavorable ones. This relabeling approach was implemented as a post-processing step, independent of the model training process, allowing us to tune fairness after calibration.

To operationalize the method, we extended our implementation to explicitly track rejected instances, distinguish them from confidently predicted samples, and apply fairness-aware relabeling accordingly. We further evaluated the impact of these changes using group-level fairness metrics including statistical parity, equal opportunity, and equalized odds. We opted not to implement the ensemble rejection (ER) or situational rejection (SR) extensions also described by Kamiran et al., as these methods require more complex ensemble logic or local similarity-based reasoning (e.g., k-nearest neighbors) and come with higher computational cost. More importantly, our focus was on evaluating the baseline probabilistic rejection (PR) method, which offers clearer interpretability and is easier to integrate into an existing pipeline. We also did not attempt to balance fairness across multiple sensitive attributes simultaneously, as the PR method is not designed for joint optimization and doing so could introduce conflicting adjustments across intersecting subgroups.

\paragraph{Post-hoc Bias Scoring Is Optimal For Fair Classification} \label{bias scoring}
We implemented the \textit{Modification with Bias Scores (MBS)} method described by Chen et al.\ \cite{chen_post-hoc_2024} to perform post-hoc fairness adjustments. For each sensitive attribute, we computed a composite bias score for each prediction by combining a group membership classifier with the model's confidence:
 
\begin{enumerate}
    \item A logistic regression model was trained to predict group membership from \texttt{calibrated\_pred\_prob}, yielding group probabilities $\mathbb{P}(A = a \mid x)$.
    \item The bias score was computed as $s(x) = (2\hat{y} - 1) \cdot (\mathbb{P}(A{=}0 \mid x) - \mathbb{P}(A{=}1 \mid x)) \,/\, (2|\texttt{prob} - 0.5| + \epsilon)$, where $\hat{y}$ is the calibrated binary prediction and $\epsilon = 10^{-8}$ prevents division by zero.
\end{enumerate}
 
We then searched over a range of bias score thresholds ($t \in [-5, 5]$, 100 steps). For each threshold $t$, predictions with $s(x) > t$ were flipped ($\hat{y} \to 1 - \hat{y}$). The optimal threshold was selected to minimize fairness disparity while maintaining accuracy above a floor.
 
We employed a cascading fairness constraint: we first sought solutions meeting disparity $\leq 0.05$ with accuracy $\geq 0.70$. If no such solution existed, the constraint was relaxed to $\leq 0.10$, then $\leq 0.15$. If no constraint was achievable above the accuracy floor, we selected the solution with the lowest disparity among those with accuracy $\geq 0.70$. This cascading approach was necessary because the feasibility of strict fairness constraints depends on the base rate gap between groups, a gap that varies substantially across demographic attributes. For each attribute, we evaluated three fairness metrics (statistical parity, equal opportunity, equalized odds) and selected the configuration achieving the highest accuracy among those meeting the applicable constraint.
 
Final results: for age, the method met $\leq 0.05$ on equal opportunity (accuracy: 0.839); for gender, $\leq 0.05$ on equal opportunity (accuracy: 0.793); for residency, the $\leq 0.05$ constraint was infeasible without destroying predictions, and the method cascaded to $\leq 0.10$ (accuracy: 0.839).

\paragraph{Bias Mitigation Post-processing for Individual and Group Fairness} \label{bias mitigation}
To address group fairness in predictive outcomes, we implemented a post-hoc bias mitigation approach inspired by the Individual and Group Debiasing (IGD) framework proposed by Lohia et al.\ \cite{lohia_bias_2018}. Their method identifies individually biased predictions through counterfactual analysis, estimating how a prediction would change if a student's demographic attribute were different, and corrects those showing the largest bias.
 
Since the vendor model cannot be queried with perturbed inputs under procurement constraints, we trained a surrogate logistic regression model on the available input features and used it to approximate counterfactual queries. For each sensitive attribute, the debiasing pipeline proceeded as follows:
 
\begin{enumerate}
    \item \textbf{Surrogate model:} A logistic regression was trained on input features to predict \texttt{SUCCESS\_LEVEL}.
    \item \textbf{Bias score computation:} For each student, we computed $\text{bias} = P_{\text{surrogate}}(\text{success} \mid \text{original}) - P_{\text{surrogate}}(\text{success} \mid \text{flipped attribute})$, comparing the surrogate's predictions with the student's actual attribute value versus a counterfactual value.
    \item \textbf{Bias detection:} Students with bias scores above the 80th percentile were flagged. A secondary logistic regression (the ``bias detector'') was trained on the unprivileged group (identified as the numerically smallest group) to predict bias flag status from input features.
    \item \textbf{Correction:} For flagged students in the unprivileged group where the bias detector predicted high bias probability ($>0.5$), predictions were replaced with the surrogate's output under the counterfactual attribute value.
\end{enumerate}
 
This pipeline was run independently for each of the three sensitive attributes (age, gender, residency status), producing separate adjusted predictions per attribute. A key design choice, identifying the ``unprivileged'' group as the numerically smallest group, proved consequential for residency status, where the smaller group (international students) already had higher positive rates than the larger group (domestic students). The method's favorable corrections were therefore directed at the group that was already advantaged on this metric, widening rather than narrowing the disparity. This misdirected correction illustrates how methods can embed normative assumptions about disadvantage that conflict with actual disparity patterns in a given dataset.

\paragraph{Group-Balanced Conformal Prediction} \label{conformal}
To assess and mitigate group-level disparities in our calibrated binary classifier, we implemented group-balanced conformal prediction (GBCP) as described in the tutorial by Angelopoulos and Bates (2022) \cite{angelopoulos_gentle_2022}. GBCP is a post-hoc thresholding technique that adjusts decision boundaries independently for each group within a sensitive attribute, aiming to align group-level positive prediction rates to a shared target. For each attribute—\texttt{age group}, \texttt{gender}, and \texttt{residency}—we estimated a threshold for each group by computing the empirical quantile of predicted probabilities such that the proportion of scores exceeding the threshold matched a specified target rate (e.g., 0.5). Formally, for a group \( g \), the threshold \( \tau_g \) was set as \( Q_{1-\alpha}(S_g) \), where \( S_g \) is the distribution of predicted probabilities in group \( g \), and \( \alpha \) is the desired target rate. Binary predictions were then generated by comparing each instance’s predicted probability to the threshold corresponding to their group membership, producing an adjusted output.

We evaluated fairness improvements in terms of true positive rate (TPR), false positive rate (FPR), and positive prediction rate (statistical parity), computing maximum disparity across groups within each attribute. To ensure reliable quantile estimates and avoid overfitting, we elected not to jointly calibrate across the full intersection of sensitive attributes, as many intersectional subgroups had insufficient support. Instead, we applied GBCP independently to each attribute, enabling isolated analysis of fairness tradeoffs across individual group axes. The target rate of 0.5 was selected to simulate a uniform decision boundary across groups and to support parity in positive classification rates, aligning with established definitions of statistical parity. This approach balances implementation simplicity, statistical robustness, and institutional interpretability, and is designed to surface fairness--accuracy tradeoffs in ways that are replicable and actionable for practitioners.

\paragraph{xOrder: Attempted Application and Exclusion} \label{towards}
We attempted to apply the \textit{xOrder} post-hoc adjustment \cite{cui_towards_2021}, which reorders predictions across groups to balance ranking fairness (measured by xAUC) against classification utility, using a dynamic program over paired group score sequences. Two obstacles led us to exclude it from our evaluation.

First, the method is defined for exactly two groups. \citet{cui_towards_2021}'s cross-group ordering takes two ordered score sequences as input; Algorithm 1 operates on those two sequences; and the fairness bound is stated in terms of the two group sizes. No generalization to $K > 2$ groups is specified, and every sensitive attribute in the original evaluation is binary. Our age attribute has four categories and our gender attribute three. Applying xOrder to these attributes would require an aggregation rule for combining pairwise comparisons into a single per-student prediction, which the method does not provide and which we would therefore have to invent.

Second, the dynamic program does not scale to institutional data. Benchmarking the algorithm on our score distributions gives empirical runtime scaling of approximately $n^{2.36}$ and memory scaling of approximately $n^{1.80}$. Extrapolating to our actual group sizes, a single pairwise comparison for residency status (24,400 domestic vs.\ 17,738 international students) would require on the order of months of computation and over one hundred gigabytes of memory. The subsampling to 100 students per group that our initial implementation used was not a design choice but a necessity, and it produces group metrics that vary depending on which comparison group is used, with no principled way to reconcile them.

We report this exclusion rather than omitting the method silently, because the reasons bear directly on our argument: a method recommended in the fairness literature for deployment-constrained settings cannot be deployed on a real institution's demographic data at a real institution's scale.
\end{document}